\documentclass[STIX1COL]{WileyNJD-v2}

\usepackage{xcolor}
\usepackage{graphicx}
\usepackage{pgfplots} 
\pgfplotsset{width=7cm,compat=1.14}
\usepgfplotslibrary{statistics}
\usepackage{tikz} 
\usetikzlibrary{calc,patterns,shapes}
\usetikzlibrary{arrows.meta}
\tikzset{>={Latex[width=2mm,length=2mm]}}
\usetikzlibrary{decorations.pathreplacing}
\usepackage{setspace}
\usepackage[font={normalsize}]{subcaption}
\usepackage{booktabs}
\usepackage{pgfplotstable}
\usepackage{microtype} 
\usepackage{listings}
\usepackage{siunitx}
\usepackage{array}

\newcolumntype{?}{!{\vrule width 1pt}}

\definecolor{mGreen}{rgb}{0,0.6,0}
\definecolor{mGray}{rgb}{0.5,0.5,0.5}
\definecolor{mPurple}{rgb}{0.58,0,0.82}
\definecolor{backgroundColour}{rgb}{0.95,0.95,0.92}

\newcommand{\bq}{\begin{equation}}
\newcommand{\eq}{\end{equation}}
\newcommand{\bytes}{\mbox{bytes}}
\newcommand{\byte}{\mbox{byte}}
\newcommand{\second}{\mbox{s}}

\newcommand{\flop}{\mbox{flop}}
\newcommand{\flops}{\mbox{flops}}
\newcommand{\NJFLOP}{\mbox{nJ/\flop}}

\newcommand{\cycle}{\mbox{cy}}
\newcommand{\iter}{\mbox{it}}
\newcommand{\cycles}{\mbox{cy}}

\newcommand{\BIT}{\mbox{\byte/\iter}}

\newcommand{\GCS}{\mbox{G\cycle/\second}}

\newcommand{\BC}{\mbox{\byte/\cycle}}
\newcommand{\GBS}{\mbox{G\byte/\second}}

\newcommand{\GFS}{\mbox{G\flop/\second}}

\newcommand{\lup}{\mbox{LUP}}

\newcommand{\GHZ}{\mbox{GHz}}

\newcommand{\FB}{\mbox{\flop/\byte}}
\newcommand{\BL}{\mbox{\byte/\lup}}
\newcommand{\GB}{\mbox{GB}}
\newcommand{\KB}{\mbox{kB}}
\newcommand{\MB}{\mbox{MB}}
\newcommand{\GiB}{\mbox{GiB}}
\newcommand{\MiB}{\mbox{MiB}}
\newcommand{\KiB}{\mbox{KiB}}
\newcommand{\W}{\mbox{W}}

\newcommand{\muops}{\mbox{$\mu$-ops}}
\newcommand{\eos}{\;.}
\newcommand{\cma}{\;,}
\newcommand{\rlm}{roof{}line model}
\newcommand{\rl}{roof{}line}

\newcommand{\ecmspace}{\,}
\newcommand{\TOL}{$T_{\mathrm{c}\_\mathrm{OL}}$}
\newcommand{\NNZR}{$N_\mathrm{nzr}$}
\newcommand{\NR}{$N_\mathrm{r}$}
\newcommand{\NNZ}{$N_\mathrm{nz}$}

\newcommand{\epsep}{\rceil}
\newcommand{\ecmp}[4]{\mbox{$\left\{{#1}\ecmspace\epsep\ecmspace {#2}\ecmspace\epsep\ecmspace {#3}\right\}\ecmspace{#4}$}}
\newcommand{\ecme}[4]{\mbox{$\left({#1}\ecmspace\epsep\ecmspace {#2}\ecmspace\epsep\ecmspace {#3}\right)\ecmspace{#4}$}}
\newcommand{\sellcs}{SELL-\texorpdfstring{$C$-$\sigma$}{C-sigma}}
\newcommand{\likwid}{\texttt{LIKWID}}
\newcommand{\likwidperfctr}{\texttt{likwid-perfctr}}
\newcommand{\likwidpin}{\texttt{likwid-pin}}

\definecolor{tumbleweed}{rgb}{0.87, 0.67, 0.53}

\newcommand{\afx}{A64FX}
\newcommand{\spmv}{SpMV}
\newcommand{\cmg}{CMG}
\newcommand{\mve}{MVE}
\newcommand{\crs}{CRS}
\newcommand{\ellpack}{ELLPACK}
\newcommand{\tands}{dRECT}

\newcommand{\Figure}{Figure}

\newcommand{\mathspace}{\text{ }}
\newcommand{\rAdd}[1]{{\color{black}{#1}\color{black}}}

\newenvironment{customlegend}[1][]{%
	\begingroup
	\csname pgfplots@init@cleared@structures\endcsname
	\pgfplotsset{#1}%
}{%
	\csname pgfplots@createlegend\endcsname
	\endgroup
}%
\def\addlegendimage{\csname pgfplots@addlegendimage\endcsname}

\definecolor{applegreen}{rgb}{0.55, 0.71, 0.0}
\definecolor{amethyst}{rgb}{0.6, 0.4, 0.8}
\definecolor{amber}{rgb}{1.0, 0.75, 0.0}

\articletype{Special Issue Paper}%

\received{}
\revised{}
\accepted{}

\begin{document}

\title{ECM modeling and performance tuning of
	SpMV and Lattice QCD on A64FX}
\author[1]{Christie Alappat*}
\author[2]{Nils Meyer}
\author[1]{Jan Laukemann}
\author[1]{Thomas Gruber}
\author[1]{Georg Hager}
\author[1]{Gerhard Wellein}
\author[2]{Tilo Wettig}

\authormark{Alappat \textsc{et al}}

\address[1]{\orgdiv{Erlangen National High Performance Computing Center}, \orgname{Friedrich-Alexander-Universität Erlangen-Nürnberg}, \orgaddress{\state{Erlangen}, \country{Germany}}}
\address[2]{\orgdiv{Department of Physics}, \orgname{University of Regensburg}, \orgaddress{\state{Regensburg}, \country{Germany}}}

\corres{*Christie Alappat,
	Erlangen National High Performance Computing Center,
	Friedrich-Alexander-Universität Erlangen-Nürnberg,
	Erlangen,
	Germany. \email{christie.alappat@fau.de}}

\abstract[Summary]{The \afx\ CPU is arguably the most powerful
  Arm-based processor design to date. Although it is a traditional
  cache-based multicore processor, its peak performance and memory
  bandwidth rival accelerator devices.  A good understanding of its
  performance features is of paramount importance for developers who
  wish to leverage its full potential.  We present an architectural
  analysis of the \afx\ used in the Fujitsu FX1000 supercomputer
  at a level of detail that allows for the
  construction of Execution-Cache-Memory (ECM) performance models for
  steady-state loops. In the process we identify architectural
  peculiarities that point to viable generic optimization
  strategies. After validating the model using simple streaming loops
  we apply the insight gained to sparse matrix-vector multiplication
  (\spmv) and the domain wall (DW) kernel from quantum chromodynamics (QCD).
  For \spmv\ we show why the \crs\ matrix storage format is not a good
  practical choice on this architecture and how the \sellcs\ format
  can achieve bandwidth saturation. For the DW kernel we provide
  a cache-reuse analysis and show how an appropriate
  choice of data layout for complex arrays can realize memory-bandwidth
   saturation in this case as well. A comparison with
  state-of-the-art high-end Intel Cascade Lake AP and Nvidia V100
  systems puts the capabilities of the \afx\ into perspective.
  We also explore the potential for power optimizations using
  the tuning knobs provided by the Fugaku system, achieving
  energy savings of about 31\% for \spmv\ and 18\% for DW.}

\keywords{ECM model, A64FX, sparse matrix-vector multiplication, lattice quantum chromodynamics}

\maketitle
\section{Introduction}\label{sec:intro}
The processor architectures used in HPC systems have been dominated for a long time by general-purpose commodity off-the-shelf processors (CPUs).
Increasing clock speeds in the past and steadily increasing core counts in the last decade
resulted in an attractive price-performance ratio at moderate power consumption.
Traditional HPC-oriented architectures such as vector computers have almost been superseded.
As power constraints and technology scaling limits became more pressing, a strong trend towards diversification in processor architectures for HPC started.
General-Purpose Graphics Processing Units (GPGPUs) provide new levels of price-performance and energy-per-flop efficiency and therefore  have become very attractive for several application fields  such as classical molecular dynamics, fluid dynamics or linear solvers as well as artificial intelligence.

Large initiatives have started to design custom HPC processors addressing the performance characteristics of a broad range of applications from computational science and engineering.
They  make use of new memory technologies or modular instruction sets and implement HPC-specific hardware concepts such as fast on-chip synchronization or specific on-chip accelerators.
The Post-K  and the European Processor Initiative (EPI)  projects are two such well-known endeavors.
The former has already delivered the Fujitsu \afx\ processor, which powers the fastest machine on the Top500 list as of November 2020, Fugaku.

The \afx\ CPU is the second design (after Intel's Xeon Phi Knights Landing) that  connects a classic cache-based multicore processor to high bandwidth memory (HBM).
While the use of HBM is established on GPGPUs with their massively threaded programming and execution model, it is an interesting question if standard CPU-oriented programming models (e.g., OpenMP) in combination with the limited thread- and data-level parallelism of the CPU hardware can also exploit the potential of HBM.
Several other features such as hardware barrier and sector cache have been implemented in the \afx\ to address the needs of HPC as well as artificial intelligence (AI) applications.
At the same time, a strict power budget had to be kept, enforcing compromises in the design of cores, caches and the chip.
Finally, the application performance of the \afx\  critically depends on the quality of its rather new software ecosystem, in particular compilers and numerical libraries.
This complex situation requires a careful analysis of existing well-optimized CPU codes, e.g., to what extent they may exploit the benefits of the new design and how the new concepts implemented in the \afx\ interfere with code-optimization techniques, parallelization strategies and data layouts.

In this paper we use analytical performance modeling and the Execution-Cache-Memory (ECM) performance  model to investigate and understand basic performance capabilities and new performance and power-saving features of the \afx\ with a focus on streaming loops.
In view of the CPU's high memory bandwidth ($> 800\,\GBS$) and moderate core count (48), the ECM model's capability to identify single-core performance contributions will be of central importance.
We choose two case studies for in-depth application performance analysis representing important fields with moderate to low computational intensities:
a sparse matrix-vector multiplication (\spmv) kernel and a Lattice QCD domain wall kernel.
Our analytical modeling approach allows us to pinpoint inefficiencies of the hardware design and the existing software ecosystem and provides recommendations on code optimization and data layouts.
We further compare the performance characteristics of the \afx\ with a high-end commodity server CPU system (Intel Cascade Lake AP) and a GPGPU (NVIDIA V100). The V100 uses a comparable HBM technology.

\paragraph{Outline} The paper is organized as follows: Section~\ref{sec:testbed} describes the basic benchmarking methodology together with the compilers and libraries used. 
It further briefly summarizes the relevant performance characteristics of the standard CPU and GPGPU systems chosen for comparison. 
A detailed architectural analysis of the A64FX-based Fujitsu FX1000 used in the Fugaku system is provided in Sec.~\ref{sec:arch}. 
Strong focus is put on the in-core analysis, including a discussion of the capabilities of Arm's Scalable Vector Extension (SVE) and the out-of-order back end. Furthermore we discuss an additional feature set of the Fugaku system: the zero fill instruction which prevents write-allocate transfers, the hardware barrier and the sector cache. 
In Sec.~\ref{sec:ecm} we establish the ECM machine model for the A64FX and validate it for a broad range of streaming kernels.
An analysis of \spmv\ performance on the A64FX is presented in Sec.~\ref{sec:spmvm}. 
Starting with a standard CPU-friendly \spmv\ data format we identify  shortcomings on the single-core level through the ECM model. 
We investigate the use of a vector-friendly data layout and  of sector cache to fully exploit the available bandwidth. 
In Sec.~\ref{sec:qcd} we address the large application field of Lattice QCD focusing on the domain wall kernel. The ECM model again guides the investigation of potential performance gains through code-optimization strategies and appropriate choices of data layout.
The impact of A64FX's power-saving mechanisms and performance comparisons  with standard CPU and GPGPU are presented in Secs.~\ref{sec:spmvm} and \ref{sec:qcd} for both case studies.
In Sec.~\ref{sec:summary} we summarize our findings and put our work in the context of existing literature. 

\paragraph{Extended version of workshop short paper}
The work presented here is an extended version of a short paper published at the PMBS 2020 workshop~\cite{PMBS20_A64FX}. The short paper investigated the basics of the ECM model and briefly demonstrated the benefit of a vector-friendly data layout for the A64FX processor used in the QPACE~4 (Fujitsu FX700)  system. 
Both topics have now been investigated on the FX1000 system used in Fugaku.
More importantly, we have substantially increased the scope of both topics, e.g., by improving the ECM model considering the impact of page sizes and by presenting a detailed ECM model and performance-tuning strategies for \spmv.
Topics presented here but not covered in~\cite{PMBS20_A64FX} include the case study of the Lattice QCD kernel, the investigation of power-saving mechanisms and specific hardware features of the A64FX and the comparison with state-of-the-art CPUs and GPGPUs.


\section{Testbed and experimental methodology}\label{sec:testbed}


\begin{table}[!tb]
	\centering
	\caption{Key specifications of the \afx\ CPU in the FX1000 system. ``$\oplus$'' represents an exclusive OR.}
	\label{table:testbed}
	\begin{tabular}{l c }
		\toprule
		Supported core frequency	& 2.0/2.2\,\GHZ   \\
		Number of CMGs 				& 4 \\
		Cores/threads per CMG       & 12/12  \\
		Instruction set         	& Armv8.2-A+SVE \\
		Max. SVE vector length  	& 512 bit\\
		Peak \flop\ rate			& 3379.2\,\GFS\\
		Cache line size         	& 256\,\bytes\\
		L1 cache capacity       	& 48$\times$64\,\KiB  \\
		L1 bandwidth per core ($b_{\mathrm{Reg}\leftrightarrow\mathrm{L1}}$)	& 128\,B/cy LD $\oplus$ 64\,B/cy ST\\
		L2 cache capacity       	& 4$\times$8\,\MiB \\
		L2 bandwidth per core ($b_{\mathrm{L1}\leftrightarrow\mathrm{L2}}$) 	&  64\,B/cy LD $\oplus$ 32\,B/cy ST \\
		Memory configuration    	& 4$\times$8\,\GiB\ HBM2 \\
		CMG \textsc{triad} bandwidth & 213\,\GBS\\
		CMG read-only bandwidth 	&  227\,\GBS\\
		Full chip \textsc{triad} bandwidth & 841\,\GBS\\
		Full chip read-only bandwidth 	&  859\,\GBS\\
		L1 translation lookaside buffer & 16 entries  \\
		L2 translation lookaside buffer & 1024 entries\\
		\bottomrule
	\end{tabular}
\end{table}

The majority of this work was done on the Fugaku supercomputer.
The system is running the Red Hat Enterprise Linux~(RHEL) 8.3 operating system.
The CPU supports two core clock frequencies, 2.0\,\GHZ\ and 2.2\,\GHZ.
The clock frequency was fixed to 2.2\,\GHZ\ unless specified otherwise.
The key specifications can be found in Table~\ref{table:testbed} and will be discussed in more detail in Sec.~\ref{sec:arch}.
All code was compiled with the GNU gcc (GCC~10.2.0) and Fujitsu~tcsds~1.2.30 (FCC~4.4.0a) compilers.
For GCC we used the \texttt{-march=armv8.2-a+sve} and \texttt{-Ofast} flags in combination with huge pages for compilation.
For FCC there exist two modes of compilation, \emph{Trad} and \emph{Clang} mode. 
We used Trad mode with  \texttt{-Kfast} and \texttt{-Khpctag} for all the runs, except for the domain wall QCD code where we used Clang mode with \texttt{-Ofast} due to the incompatibility of Trad mode with GCC attributes.
Possible deviations from the default compiler flags stated above will be mentioned in the relevant sections.

All benchmarks were run in double precision so that a vector length (VL) of 512 bits corresponds to eight real or four complex elements.
In general, SVE vector intrinsics (ACLE~\cite{ACLE}) were employed to have better control over code generation.
All arrays were aligned to OS page boundaries, for which best performance was observed in the experiments.
In order to minimize statistical variations we repeated every benchmark loop for an overall runtime
of at least one second.
We do not show the statistical fluctuations if they were below 5\%.

For benchmarking individual machine instructions we employed the ibench~\cite{ibench} framework and cross-checked our results with the \afx\
Microarchitecture Manual~\cite{a64fx_manual}.
The \likwid~\cite{Treibig:2010:2} tool suite in version 
v5.1.0~\cite{gruber_thomas_2020_4282696} was used, specifically \likwidperfctr\ for counting
hardware events and \likwidpin\ to pin the threads to cores.
The Power API framework~\cite{PowerAPI}~v2.0 installed on Fugaku was used for setting the power tuning knobs and measuring the energy consumption.

To put the results in relation to state-of-the-art hardware currently available, experiments were also run on an Intel Cascade Lake (CLX-AP) and an NVIDIA V100 GPU system.
The CLX-AP experiments were conducted on a dual-socket Intel Xeon Platinum~9242 node with 
96 cores and a STREAM \textsc{triad} bandwidth of 464\,\GBS.
For compilation the Intel compiler version 19.1.3 was used.
The GPU experiments were performed on an NVIDIA Tesla V100  connected via  PCI-Express.
The GPU kernels were compiled using CUDA~v10.2 \rAdd{with GCC~8.1.0 as host compiler}.
The V100 GPU achieves a STREAM \textsc{triad} bandwidth of 840\,\GBS, similar to that of \afx.

\section{Architectural analysis}\label{sec:arch}
\subsection{In-core} \label{ssec:incore}
\begin{figure}[tb]
	\centering
	\includegraphics*[width=0.6\linewidth]{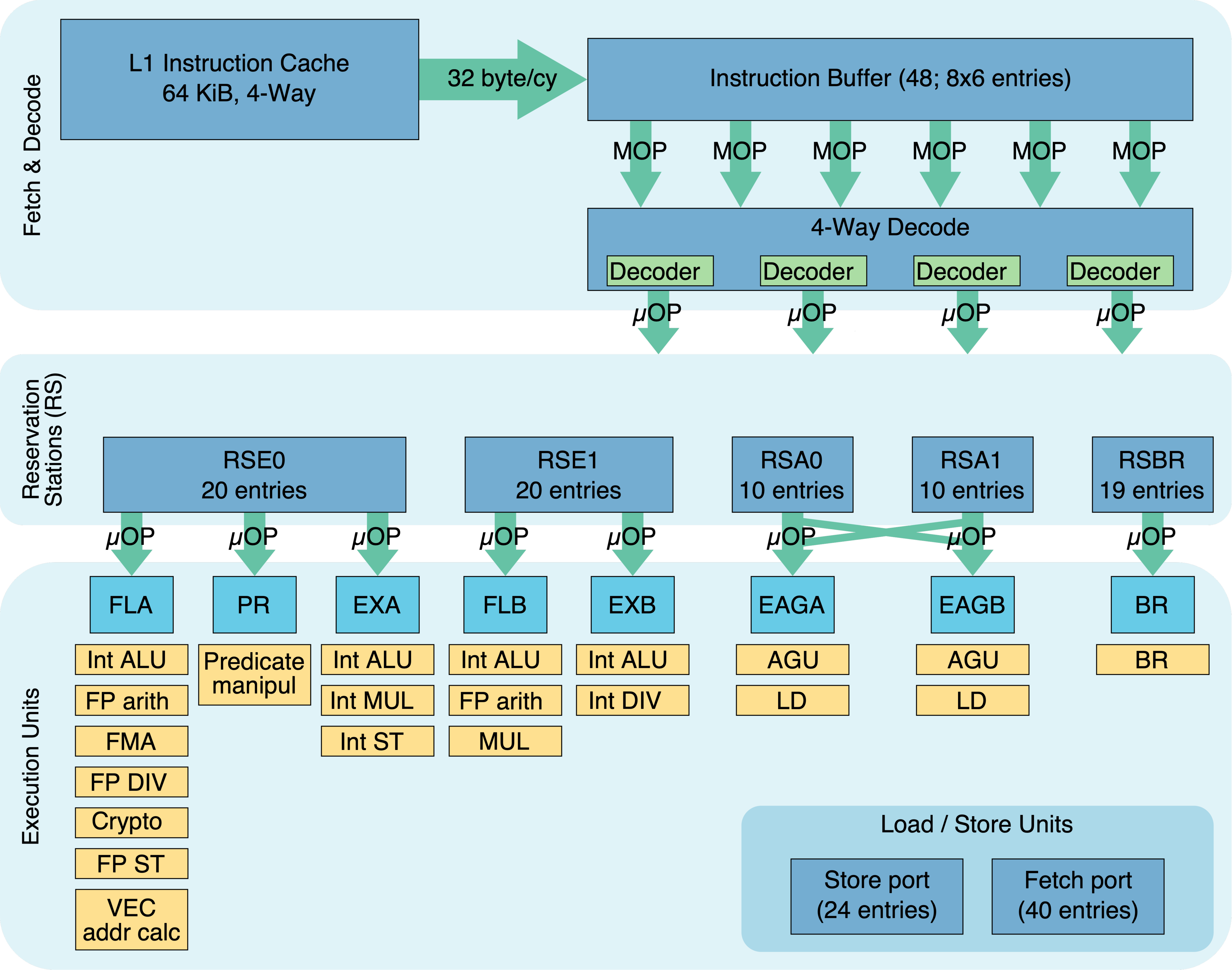}
	\caption{Overview of the in-order front end (\emph{Fetch \& Decode}) and out-of-order back end (\emph{Reservation Stations} (RS) and \emph{Execution Units}) components of an \afx\ core based on~\cite{a64fx_manual}.}
	\label{fig:a64fx_uarch}
\end{figure}
To get a better understanding of the in-core behavior of the \afx\ micro\-archi\-tecture, a schematic block diagram of one core's front-end and back-end components is shown in Fig.~\ref{fig:a64fx_uarch}.
Instructions are fetched from the L1 instruction cache with a bandwidth of 32\,\BC.
Each core's front end has an instruction buffer with $6\times8$ entries, which can feed six instructions, also called \emph{macro-operations} (MOP), per cycle to the decoder.
The decoder feeds up to four micro-operations (\muops) per cycle to the different reservation stations~(RS), which then schedule the \muops\ to the corresponding execution pipelines.
The reservation stations RSA0 and RSA1 are used for address generation and load units and have ten entries each, while reservation stations RSE0 and RSE1 are used for arithmetic execution and store units and have 20 entries each.
The reservation station RSBR with its corresponding pipeline is only used for branching and has 19 entries.
Excerpts of the execution units of the residual pipelines FL[A|B], PR, EX[A|B], EAG[A|B] are shown in boxes below the pipeline name.
While RSA0 and RSA1 can schedule instructions to both EAG pipelines, all other RS are limited to their corresponding pipelines, even if an execution unit with equivalent functionality exists in a pipeline outside their scope.
Load (LD) and store (ST) instructions are fed to the fetch port and store port, respectively, which execute the requests in parallel to the operation flow.

The small capacity of the reservation stations in combination with high instruction latencies can result in inefficient out-of-order~(OoO) execution, which emphasizes the importance of a compiler that is capable of exploiting the in-core performance by intelligent code generation.
While these hardware constraints cannot be overcome completely, their impact can be alleviated by techniques such as loop unrolling, consecutive addressing and interleaving of different instruction types. These techniques have been beneficial in our benchmarks, see Secs.~\ref{sec:ecm} and \ref{sec:spmvm} for details.
Neither the open-source GCC compiler (version 10.2.0) nor Fujitsu's proprietary FCC compiler (version 4.4.0a) currently generate code that fully overcomes the hardware constraints, which is why we employed SVE intrinsics for our benchmarks.

To create an accurate in-core model of the \afx\ micro\-archi\-tecture, we 
analyze different \emph{instruction forms}, i.e., assembly instructions in 
combination with their operand types, based on the methodology introduced 
in~\cite{OSACA2018, OSACA2019}.\footnote{See 
\href{https://github.com/RRZE-HPC/OSACA}{github.com/RRZE-HPC/OSACA} for a full 
set of measured instruction forms and the \afx\ port model used in this work.}
Table~\ref{table:incore-instructions} shows a list of instruction forms relevant for this work.
Standard SVE load (\texttt{ld1d}) instructions have a reciprocal throughput of 0.5\,cycle~(\cycle), while stores (\texttt{st1d}) have 1\,\cycle.
The throughput of gather instructions depends on the distribution of addresses:
``simple'' access patterns are stride 0~(no stride), 1~(consecutive load) and 2, while larger strides and irregular patterns are considered ``complex.'' The former have lower reciprocal throughput and latency than the latter.
However, when occurring in combination with a standard LD for loading the index array,
we can observe an increase of reciprocal throughput by 1.5\,\cycles\
instead of the expected 0.5\,\cycle.
This is caused by the
dependency of the gather instruction on the preceding index load operation, which the OoO execution
cannot hide completely.
Note also the rather long latencies for arithmetic operations such as MUL, ADD and FMA compared to other state-of-the-art architectures (e.g., on Intel Skylake or AMD Zen2 these are between 3\,\cycles\
and 5\,\cycles).

The SVE instruction set introduced a ``\texttt{while}\{cond\}'' instruction to set predicate registers
according to the elements in vector registers in a length-agnostic way
in order to eliminate remainder loops.
Although it is used extensively for SVE code, a port-conflict analysis revealed that this instruction does not collide with floating-point instructions or data transfers.

\begin{table}[!tb]
	\centering
	\caption{In-core instruction throughput and latency (if applicable) for selected instruction forms.}
	\label{table:incore-instructions}
	\begin{tabular}{l c c}
		\toprule
		\multirow{2}{*}{\textbf{Instruction}} & \textbf{Reciprocal} & \multirow{2}{*}{\textbf{Latency [cy]}} \\
		& \textbf{Throughput [cy]} & \\
		\hline
		\texttt{ld1d} (standard)	& 0.5 & 11 \\
		\texttt{ld1d} (gather, simple stride)& 2.0 & $\geq$ 11 \\
		\texttt{ld1d} (gather, complex stride)& 4.0 & $\geq$ 11 \\
		simple gather + standard load			& 3.5 & -- \\ 
		complex gather + standard load			& 5.5 & -- \\ 
		\texttt{st1d} (standard)			& 1.0 & -- \\ 
		\texttt{fadd}                   & 0.5 & 9 \\
		\texttt{fmad}                   & 0.5 & 9 \\
		\texttt{fmla}                   & 0.5 & 9 \\
		\texttt{fmul}                   & 0.5 & 9 \\
		\texttt{fcadd}                  & 1.0 & 15 \\
		\texttt{fcmla}                  & 2.0 & 16 \\
		\texttt{fadda} (512 bit) & 18.5 & 72 \\
		\texttt{faddv} (512 bit) & 11.5 & 49 \\
		\texttt{while\{le|lo|ls|lt\}}                   & 1.0 & 1 \\
		\bottomrule
	\end{tabular}
\end{table}

\subsection{Chip topology and memory hierarchy} \label{ssec:memhier}
The chip is divided into four \emph{core memory groups} (CMG) of twelve cores each.
Every CMG is its own  \rAdd{cache-coherent non-uniform memory access} (ccNUMA) domain. The 64\,\KiB\ L1 cache is \rAdd{core-local},
while 8\,\MiB\ of L2 are shared among the cores of a CMG.
We refer to Table~\ref{table:testbed} for architectural details of the A64FX processor in the Fugaku system.

While parallel load/store from and to L1 cache is possible
for general-purpose and NEON registers, different types of
SVE data-transfer instructions in L1 cannot be executed in
one cycle: Using SVE, one \afx\ core can either load up to
$2 \times 64$\,\BC\ or store $64$\,\BC\ from/to L1. The L2 cache
can deliver $64$\,\BC\ to one L1 but tops out at $512$\,\BC\ per
CMG.
The L1-L2 write bandwidth is half the load bandwidth,
i.e., $32$\,\BC\ per core, and is capped at $256$\,\BC\ per CMG.
Finally, the measured main-memory bandwidth  per CMG is $227$\,\GBS\ for read-only and $213$\,\GBS\ for STREAM \textsc{triad}.
The memory bandwidth scales almost linearly with the CMGs and reaches a full-chip read-only bandwidth of $859$\,\GBS\ and STREAM \textsc{triad} bandwidth of $841$\,\GBS.
 These measured bandwidths will be used as baselines for the memory-transfer bandwidth in the ECM model.

\subsection{Special features of A64FX}
The \afx\ processor has some special hardware capabilities to improve the performance of some codes.
In this section we look into three features, i.e., zero fill, hardware barrier and sector cache.
Currently only the FCC compiler supports these features on high-level code.

\begin{figure}[tb]
	\begin{minipage}{0.46\textwidth}
		%
	\input{plots/tikz/cl_zero.tex}%

		\caption{Reported STREAM \textsc{triad} bandwidth
		 with and without zero fill. The data-set size was $3$\,\GB.}
		\label{fig:cl_zero}
	\end{minipage}
\hspace{1em}
	\begin{minipage}{0.5\textwidth}
        \vspace{1.5em}
		%
	\input{plots/tikz/barrier.tex}%

		\caption{Comparison of the costs of different barrier implementations. GCC, FCC and FCC-active refer to software barriers. The latter
		uses a spin waiting loop.
		FCC-hard uses the hardware barrier implemented on the \afx.}
		\label{fig:hw_barrier}
	\end{minipage}
\end{figure}

\subsubsection{Zero fill}
A cache write miss causes a write-allocate transfer, i.e., the cache line must be read before it can be modified.
However, this increases the memory traffic, thus reducing the effective bandwidth available to the application.
Most processors therefore have a mechanism to
 avoid this additional traffic.
On the \afx\ processor a similar mechanism exists and is called zero fill.
The zero fill instruction \texttt{DC ZVA} directly writes
a cache line filled with zeros to the L2 cache.
Therefore, the processor can load the cache line from L2 through L1 avoiding the read operation from main memory.
This effectively increases the measured application bandwidth.

For simple codes the  FCC compiler is capable of automatically detecting the arrays which only have write operations and will use the \texttt{DC ZVA} instruction.
In order to enable this the \texttt{-Kzfill}  compiler
flag has to be used.
Figure~\ref{fig:cl_zero} shows the increase in application bandwidth when using zero fill for the STREAM \textsc{triad} (\verb.a[i]=b[i]+s*c[i].)  benchmark.
The benchmark reads two double-precision arrays and writes to one array.
The bandwidth reported by the benchmark assumes
24\,\bytes\ of data traffic per iteration (\BIT), which in reality is obtained only if the write-allocate operation is avoided.
Without zero fill there would be an additional read operation costing an extra 8\,\BIT, and thereby we only observe 3/4~th of the actual bandwidth.
This difference can be seen in the figure.

\subsubsection{Hardware barrier}
Another feature of the \afx\ processor is the hardware barrier.
The hardware barrier allows for fast synchronization of threads using dedicated system registers.
With the FCC compiler the hardware barrier can be
activated by setting the environment flag \texttt{FLIB\_BARRIER} to \texttt{HARD}.

To determine the cost of the barrier we measured the difference
in time between two variants of a  computationally intensive kernel (calculating the exponential of a number), one with \texttt{omp barrier} and the other without.
Figure~\ref{fig:hw_barrier}  compares the cost (in cycles) of hardware barrier and different types of software barriers for varying number of threads.
The cost of the default software barrier used by the GCC and FCC compilers is shown in blue and red, respectively. We see that FCC has
a small advantage here.
However, when we direct FCC's software barrier to use
a spin waiting loop by setting the environment variable \texttt{OMP\_WAIT\_POLICY} to \texttt{active}, we see that
the cost of the barrier drops further by $20$\%.
The cost of GCC's software barrier did not change by setting this environment variable.
We see that the hardware barrier performs best and requires only $550$\,\cycles\ within one CMG.
Going beyond the CMG (12 cores) the cost increases to almost $2200$\,\cycles.
This is because the hardware barrier is only implemented within a CMG, while between CMGs a software barrier is used.
Note that the results shown here are the statistics from
ten runs as the run-to-run fluctuations in this experiment were higher than 5\% for some cases.

\begin{figure}[tb]
	 \begin{minipage}[b]{0.68\textwidth}
	\begin{subfigure}[b]{0.5\textwidth}
	%
	\input{plots/tikz/sector_cache_perf.tex}%

	\caption{Performance.} \label{fig:sector_cache_perf}
	\end{subfigure}
	\hspace{-0.4em}
	\begin{subfigure}[b]{0.5\textwidth}
	%
	\input{plots/tikz/sector_cache_mem.tex}%

	\caption{Main-memory traffic.} \label{fig:sector_cache_mem}
	\end{subfigure}
	\caption{Performance and main-memory data traffic of
	dense matrix-vector multiplication for different vector sizes using sector cache.\label{fig:sector_cache}}
	\end{minipage}
\hfill
\begin{minipage}[b]{0.29\textwidth}
	  %
	\input{plots/tikz/sector_cache_legend.tex}%

	  \vspace{-5.32em}
	\begin{lstlisting}
	//loop over rows
	for(int i = 0; i < N_r; ++i)
	{
		//loop over columns
		for(int j = 0; j < N_c; ++j)
		{
			y[j] += A[i][j] * x[i];
		}
	}
	\end{lstlisting}
	\vspace{1.2em}
\captionof{lstlisting}{\rAdd{Code snippet of DAXPY-style DMVM.}}\label{listing:dmvm}
\end{minipage}
\end{figure}

\subsubsection{Sector cache}
\label{sec:sector-cache}
Sector cache is a mechanism to partition a cache into different sectors of varying size. 
The application can then tag its data structures to be directed to one of the sectors. 
This allows for more control of the cache space allocated for each data structure in the code. 
For example, in case of a code with reuse on one array and streaming patterns on other arrays, one could direct the 
streaming arrays to a small sector of the cache to avoid 
polluting the cache space that could be used by the array having reuse.
The sector size can be controlled with a granularity of cache ways.
With the FCC compiler, pragma directives are used to activate the sector cache and to tag the arrays.

Figure~\ref{fig:sector_cache} shows the impact of sector cache on a DAXPY-style dense matrix-vector multiplication (DMVM), for which the inner loop traverses a column 
of the matrix \rAdd{(see Listing~\ref{listing:dmvm})}. 
The DMVM kernel 
performs a multiplication of matrix $A$, stored in column-major format,
with vector $x$ and writes the result into vector $y$.
The matrix array $A$ does not have any reuse and therefore can be directed to one small sector of the cache.
On the other hand, the vector array $x$ is reused all the time, and the vector array $y$ is reused if its size is small enough to fit in the cache. 
In the experiment we fix the number of rows to $192$ and vary the number of columns, i.e., $x$ is fixed to length $192$ and the length of $y$ is variable.
In Fig.~\ref{fig:sector_cache_perf} we plot the performance as a function of the size of $y$.
The different lines in the figure correspond to the different sizes of the cache sector allocated for matrix array $A$.
The black line corresponds to the case without any use of sector cache. 
It can be seen that as the size of the vector $y$ increases to about $2$\,\MB\ the performance starts to drop as the vector $y$ can no longer be kept in L2. 
The drop in performance can be correlated with an additional memory data traffic of $16$\,\bytes\ (see Fig.~\ref{fig:sector_cache_mem}) due to the read and write of vector $y$.
However, if we restrict the space available to matrix $A$,
the vector $y$ has more cache space available, and therefore the kernel can sustain the high performance level until almost $5$\,\MB.
Restricting the cache available to matrix $A$ to a very small size of $1$ cache way is, however, not optimal since this
leads to an early eviction of the prefetched elements of the matrix $A$.
Obviously, the performance is also worse if we allocate almost all cache space (12 out of 14 allocatable ways) of L2 to the matrix $A$.
Note that due to the high cost of the initial invocation of the sector cache (almost 500\,milliseconds) the first call to the DMVM kernel is not included in the performance results.

\section{Construction of the ECM model}
\label{sec:ecm}
Given the information about in-core execution and data traffic across all  data paths in the memory hierarchy gathered in Secs.~\ref{ssec:incore} and \ref{ssec:memhier}, a performance model for the \afx\ can be constructed.
The Execution-Cache-Memory~(ECM) model is an analytical performance model for streaming loop kernels with regular data-access patterns and equal amount of work per loop iteration, using first principles and machine-dependent constraints. 
\rAdd{As opposed to the \rlm, the ECM model can identify
execution and data transfer bottlenecks for  single-threaded programs and predict the scaling behavior of loops across the cores of a multicore chip.
It also allows one to take into account overlapping 
or non-overlapping data transfers within the cache hierarchy. 
The \rlm\ always assumes full overlap of all
time contributions.
}

\subsection{Time contributions}
\label{subsec:ecm_contrib}
The ECM model of the \afx\ contains four different time contributions:
\begin{enumerate}
	\item \TOL: execution time for in-core instructions  that can overlap with data transfers.
	\rAdd{These are all instructions except loads.
	This also includes the cycles generated by the store instructions on the FLA and EXA pipelines.}
	\item \(T_{\mathrm{L1\_LD}}\): time for in-core load data traffic between registers and L1 cache.
	\item \(T_{\mathrm{L1\_ST}}\): time for in-core store data traffic between registers and L1 cache.
	\item Data transfer time between any other memory hierarchy levels: \(T_{\mathrm{L2}}\) for data between L1 and L2, and \(T_{\mathrm{Mem}}\) for data between L2 and main memory.
\end{enumerate}
Combining these contributions, the single-core runtime prediction for the \afx\ is defined as
\begin{align}
	T_{\mathrm{ECM}} &= \max\big(T_{\mathrm{c\_OL}}, f(T_{\mathrm{L1\_LD}}, T_{\mathrm{L1\_ST}}, T_{\mathrm{L2}}, T_{\mathrm{Mem}})\big) \cma
	\label{eq:ecm_contrib}
\end{align}
where $f$ is a combination of sum and max operators depending on the overlap hypothesis, which will be discussed in Sec.~\ref{subsec:overlap_hypo}.
For $T_i$ with $i\in$\{c\_OL, L1\_LD, L1\_ST\} the time contributions are determined by a static analysis of the assembly code using the OSACA tool.
For $T_i$ with $i\in$\{L2, Mem\} the time contributions are given by
\begin{align}
T_i &=  V_i/b_i\cma \label{eq:ecm_contrib_ti} 
\end{align}
where $V_i$ is the data volume transferred and $b_i$ is the bandwidth between memory hierarchy level $i$ and the next lower 
level.\footnote{Lower means closer to the cores, e.g.,  L1 is lower than L2.}
The $V_i$ include write-allocate transfers due to store misses where applicable.
Latency effects are neglected.

\subsection{Overlap hypothesis}
\label{subsec:overlap_hypo}
Depending on the architecture the data transfer
	contributions may or may not overlap, i.e., the function $f$ in Eq.~\eqref{eq:ecm_contrib} has to be determined.
In order to find out which of the time contributions for data transfers through the cache hierarchy overlap,  measurements for a test kernel are compared with predictions based on different hypotheses, see~\cite{JSFI310} for an in-depth description of this iterative process.
If a hypothesis works for the test kernel, it is tested against a collection of other kernels with different characteristics to validate or invalidate the hypothesis.

\begin{figure}[tb]
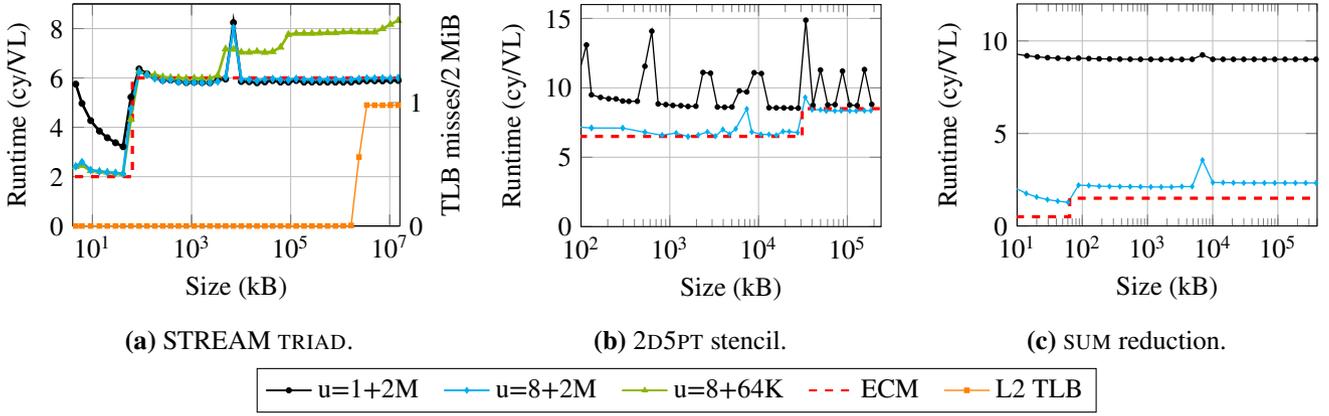

	\begin{subfigure}[tb]{0.33\textwidth}
		\centering
		%
	\input{plots/tikz/ecm/ecm_stream.tex}%

		\caption{STREAM \textsc{triad}.}
		\label{fig:ecm_stream}
	\end{subfigure}
	\hspace{0.15em}
	\begin{subfigure}[tb]{0.31\textwidth}
		\centering
        \vspace{-0.1em}
		%
	\input{plots/tikz/ecm/ecm_stencil.tex}%

		\caption{\textsc{2d5pt} stencil.}
		\label{fig:ecm_stencil}
	\end{subfigure}
	\hspace{0.1em}
	\begin{subfigure}[tb]{0.31\textwidth}
		\centering
		%
	\input{plots/tikz/ecm/ecm_sum.tex}%

		\caption{\textsc{sum} reduction.}
		\label{fig:ecm_sum}
	\end{subfigure}
	\centering
        \smallskip
        
	%
	\input{plots/tikz/ecm/ecm_stream_legend.tex}%

	\caption{Runtime of SVE loop kernels vs. problem size, comparing no unrolling (black) and eight-way unrolling (blue).
		Both versions are using 2\,\MiB\ huge pages. Arrays were aligned to 1024-\byte\ boundaries.
		While huge pages were used by default, the extra green line denotes the usage of standard 64\,\KiB\ pages.
		The orange line in (a) shows the TLB misses on 2\,\MiB\ pages. 
	For the \textsc{2d5pt} stencil the outer and inner dimension was set at a ratio of 1:2.}
	\label{fig:stream_ecm}
\end{figure}

Here we use the STREAM \textsc{triad} kernel, \verb.a[i]=b[i]+s*c[i]., to narrow down the possible overlap scenarios. This kernel has two LD, one ST and one FMA instruction per SVE-vectorized iteration, which corresponds to eight high-level iterations.
\Figure~\ref{fig:ecm_stream} shows performance in cycles per VL  for different code variants:
``u=1'' denotes no unrolling (apart from SVE) and ``u=8'' is eight-way unrolled on top of SVE.
Some level of manual unrolling (typically eight-way) is usually required for best in-core performance.
This is even more important in kernels where dependencies cannot be resolved easily by the out-of-order logic.
Measurements of the STREAM \textsc{triad} kernel with 64\,\KiB\ pages show performance degradation starting at 64\,\MiB\ due to TLB (translation lookaside buffer)  misses after all $1024$ entries of the L2 data TLB are used.
The Fujitsu compiler suite uses 2\,\MiB\ large pages (also called \emph{huge pages}) by default (\texttt{-Klargepage} option).
For other compilers, the software environment on the Fugaku system provides the \textsc{libmpg} library and a custom linker script.
TLB measurements of the STREAM \textsc{triad} kernel with 2\,\MiB\ pages in Fig.~\ref{fig:ecm_stream} show a rise in TLB misses when the working set size exceeds 2\,\GiB.
Despite the occurrence of TLB misses, there is no observable drop in performance.
For all further measurements in this work, a page size of $2$\,\MiB\ is used.

\begin{figure}[tb]
	\centering
	\includegraphics*[width=0.99\linewidth]{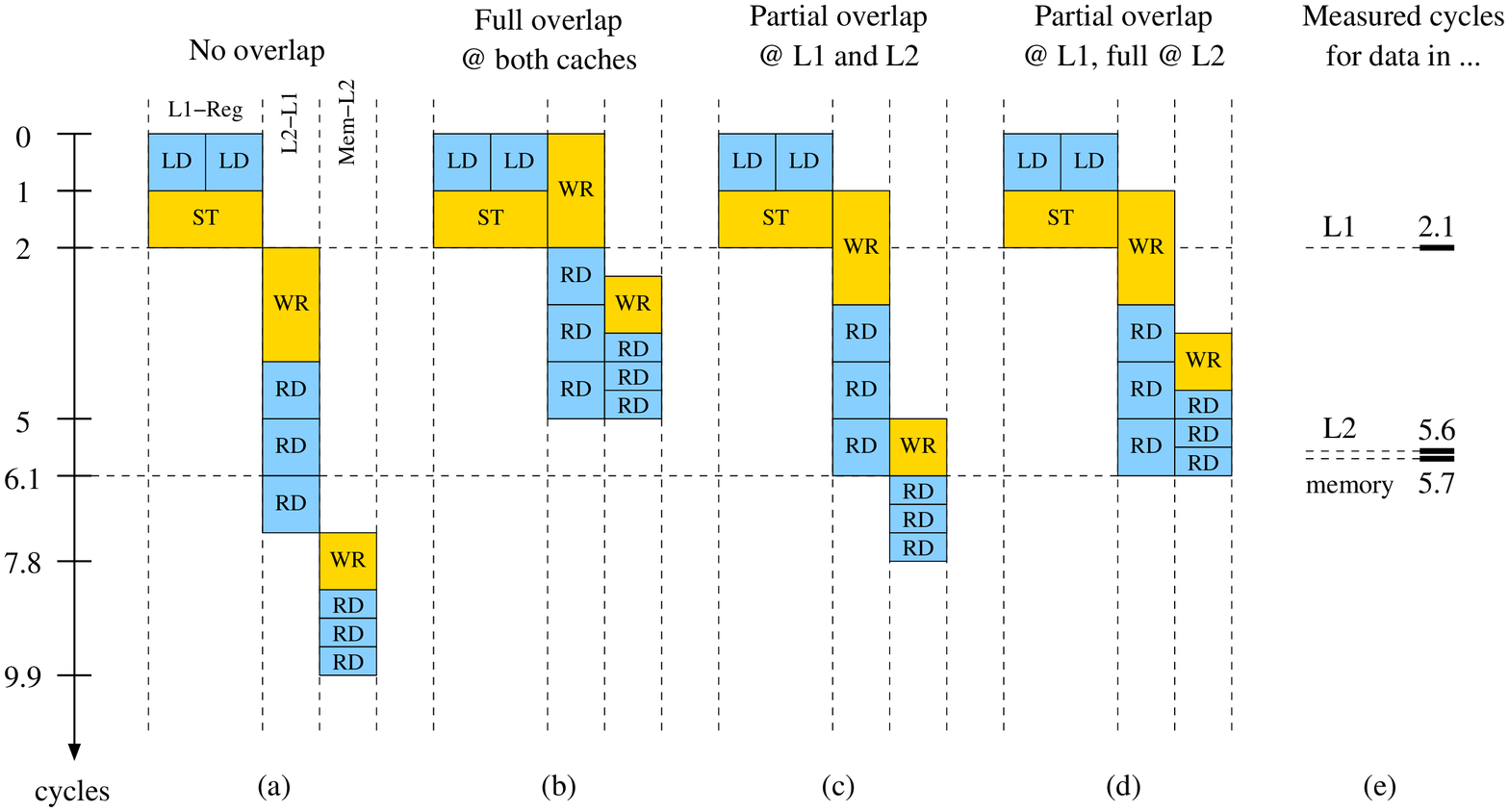}
	\caption{Comparing different overlap scenarios (a), (b), (c) and (d)
		for data transfers in the memory hierarchy with measured cycles
		per VL (e) on the STREAM \textsc{triad} kernel. Note that (c) is the
                appropriate overlap hypothesis for standard (64\,\KiB) pages.\label{fig:triad_ovl}
       }
\end{figure}

\Figure~\ref{fig:triad_ovl} compares four overlap scenarios (a), (b), (c) and (d) with measured cycles per VL (e).
Note that there is a large number of possible overlap hypotheses, and we can only show a few here.
The one leading to the best match with the STREAM \textsc{triad} data (shown in Fig.~\ref{fig:triad_ovl}d) is the following:
\begin{itemize}
	\item L1 is \emph{partially} overlapping:
	Cycles in which STs are retired in the core can overlap with L1-L2 (or L2-L1) transfers, but cycles with LDs retiring cannot.
	\item L2 is \emph{fully} overlapping:
	Cycles in which the memory interface reads and writes data from and to memory can entirely overlap with transfers between L2 and L1.
\end{itemize}
The overlap hypothesis (d) implies that the function $f$ in Eq.~\eqref{eq:ecm_contrib} has the form
\begin{align}
	f(T_{\mathrm{L1\_LD}}, T_{\mathrm{L1\_ST}}, T_{\mathrm{L2}}, T_{\mathrm{Mem}}) &= \max\big( T_{\mathrm{L1\_LD}} + \max(T_{\mathrm{L1\_ST}}, T_{\mathrm{L2}}), T_{\mathrm{Mem}}\big) \eos \label{eq:ecm_f_ol}
\end{align}
The overlap hypothesis (c) was used in previous work~\cite{PMBS20_A64FX} and applies for standard $64$\,\KiB\ pages.

In Fig.~\ref{fig:ecm_stencil} we show performance data and ECM model for a 2d five-point stencil, where SVE vectorization alone (without further unrolling) seems unable to resolve dependencies via OoO execution, leading to up to 2$\times$ slower code than the eight-way unrolled kernel despite the lack of loop-carried dependencies.
\Figure~\ref{fig:ecm_sum} shows performance data and ECM model for a sum reduction, \verb.s+=a[i]., which requires eight-way modulo variable expansion (MVE, see Sec.~\ref{subsec:spmv_mve}) on top of SVE in order to hide the large latency of the floating-point ADD instruction.

\subsection{Multicore}
For multiple cores within a \cmg, the naive scaling hypothesis of the ECM model assumes perfect bandwidth scaling across all cores of the \cmg\ until the main memory bandwidth bottleneck $b_\mathrm{Mem,\,CMG}$ of the CMG is hit.
If only a single core is active the memory bandwidth attained by a certain application is\footnote{We use lowercase characters for architectural parameters and uppercase for observed or predicted quantities.}
\begin{align}
	B(1) &= V_\mathrm{Mem}/T_\mathrm{ECM} \eos  \label{eq:single_core_bw}   
\end{align}
Assuming linear scaling of the memory bandwidth with the number of active cores, the number  $N_s$ of cores required to attain the full bandwidth $b_\mathrm{Mem,\,CMG}$ of one CMG is thus
\begin{align}
	N_s &= \lceil b_\mathrm{Mem,\,CMG}/B(1) \rceil \eos \label{eq:ecm_sat}
\end{align}
Therefore, the ECM prediction can be extended to a function of a variable number $n$ of cores,
\begin{align}
	T_{\mathrm{ECM}}(n) &= T_{\mathrm{ECM}}(1)/\min(n, N_s) \eos \label{eq:ecm_multi_core}
\end{align}
For $b_\mathrm{Mem,\,CMG}$ we use the read-only bandwidth of 227\,\GBS\ if the application is dominated by loads
and the STREAM \textsc{triad} bandwidth of 213\,\GBS\ otherwise.
If the application cannot saturate $b_\mathrm{Mem,\,CMG}$, $N_s$ will be larger than 12.

Going beyond a single CMG we again assume linear scaling since each CMG constitutes its 
own ccNUMA domain and each domain is connected to its own HBM stack.
The full-chip memory bandwidths $b_\mathrm{Mem}$ for read-only and \textsc{triad} are given in Table~\ref{table:testbed}.

\subsection{Validation of the ECM model for streaming kernels}
\label{subsec:validation}

\begin{table}
	\centering
	\caption{ECM model predictions and measurements in [cy/VL] for different streaming kernels on a single core.
		Red color indicates a deviation from the model of at least 15\%.
		The optimal unrolling factor for each measurement is shown as a subscript.}
	\label{tab:validate}
		\begin{tabular}{lll}
			\toprule
			\textbf{Kernel}  &  \textbf{Predictions} & \textbf{Measurements} \\[0.5mm]
			\midrule
			\textsc{copy} (\texttt{a[i]=b[i]}) & \ecmp{1.5}{4.5}{4.6}{} &  \ecme{1.6_{12}}{4.4_{13}}{4.6_{2}}{}\\[0.5mm]
			\textsc{daxpy} (\texttt{y[i]=a[i]*x+y[i]})  &\ecmp{2.0}{5.0}{5.1}{} &  \ecme{2.1_{8}\mathspace}{4.7_{16}}{4.7_{12}}{}\\[0.5mm]
			\textsc{dot}  (\texttt{sum+=a[i]*b[i]}) & \ecmp{1.0}{3.0}{3.1}{} &  \ecme{\textcolor{red}{1.7}_{8}\mathspace}{3.2_{4}\mathspace\mathspace}{3.3_{3}}{}\\[0.5mm]
			\textsc{init}  (\texttt{a[i]=s}) & \ecmp{1.0}{3.0}{3.1}{} &  \ecme{1.0_{13}}{2.9_{13}}{\textcolor{red}{4.0_{7}}}{}\\[0.5mm]
			\textsc{init4}  (\texttt{a[i]=s}) & \ecmp{4.0}{12.0}{12.3}{} & \ecme{4.1_{2}\mathspace}{10.6_{16}}{10.6_{16}}{}\\[0.5mm]
			\textsc{load}  (\texttt{load(a[i])}) & \ecmp{0.5}{1.5}{1.5}{} &  \ecme{\textcolor{red}{0.7}_{10}}{\textcolor{red}{2.3}_{4}\mathspace}{1.5_{1}}{}\\[0.5mm]
			\textsc{load4}  (\texttt{load(a[i])}) & \ecmp{2.0}{6.0}{6.1}{} &  \ecme{\textcolor{red}{2.5}_{4}\mathspace}{5.8_{16}}{5.9_{1}}{}\\[0.5mm]
			\textsc{triad}  (\texttt{a[i]=b[i]+s*c[i]}) & \ecmp{2.0}{6.0}{6.1}{} &  \ecme{2.1_{8}\mathspace}{5.6_{11}}{5.7_{1}}{}\\[0.5mm]
			\textsc{sum}  (\texttt{sum+=a[i]}) & \ecmp{0.5}{1.5}{1.5}{} &  \ecme{\textcolor{red}{1.1}_{11}}{\textcolor{red}{2.0}_{15}}{\textcolor{red}{2.3}_{9}}{}\\[0.5mm]
			\textsc{sch\"{o}nauer}  (\texttt{a[i]=b[i]+c[i]*d[i]}) & \ecmp{2.5}{7.5}{7.7}{} &  \ecme{2.6_{14}}{7.0_{4}\mathspace}{7.0_{4}}{}\\[0.5mm]
			\textsc{2d5pt} - LC satisfied  &  \ecmp{3.5}{6.5}{6.5}{} &  \ecme{\textcolor{red}{5.8}_{10}}{6.5_{6}\mathspace}{6.5_{9}}{}\\[0.5mm]
			\textsc{2d5pt} - LC violated in L1  &  \ecmp{3.5}{8.5}{8.7}{} &  \ecme{\textcolor{red}{5.8}_{10}}{8.6_{7}\mathspace}{8.4_{8}}{}\\[0.5mm]
			\textsc{2d5pt} - LC violated  &  \ecmp{3.5}{8.5}{8.7}{} &  \ecme{\textcolor{red}{5.8}_{10}}{8.6_{7}\mathspace}{8.4_{8}}{}\\[0.5mm]
			\bottomrule
		\end{tabular}
	\vspace{-0.8em}
\end{table}

With the in-core and data-transfer models in place we can now test the ECM model against a variety of loop kernels.
Table~\ref{tab:validate} shows a comparison of predictions and measurements.
For each kernel, three numbers represent the cycles per VL with the data set in L1, L2 and memory, respectively.
\textsc{init4} and \textsc{load4} are versions of \textsc{init} and \textsc{load} with four independent data streams.
In case of the 2d five-point stencil, three cases are shown: layer conditions~(LC) satisfied at L1, broken at L1 and broken at L2.\footnote{See~\cite{sthw15} for a comprehensive coverage of layer conditions in the context of the ECM model and Sec.~\ref{subsec:qcd_lc} below.}

The results have been obtained by running each kernel with unrolling factors from 1 to 16 and taking the best result.
Entries in red color have a deviation from the model of 15\% or more.
The strongest deviations occur in L1: Even with massive MVE, \textsc{sum} cannot achieve the architectural limit of 0.5\,\cycles/VL.
A similar deviation can be observed for the stencil kernels.
We attribute this failure to insufficient OoO resources: A modified stencil code without intra-iteration register dependencies achieves a performance within 10\% of the prediction.
Deviations from the model with L2 and memory working sets occur mainly with kernels that have a single data stream.
In fact, we can observe that the $\geq15$\% deviation for both \textsc{load} and \textsc{init} in L2 and memory, respectively, decreases to 3\% and 14\% when using four streams.

\begin{figure}[tb]
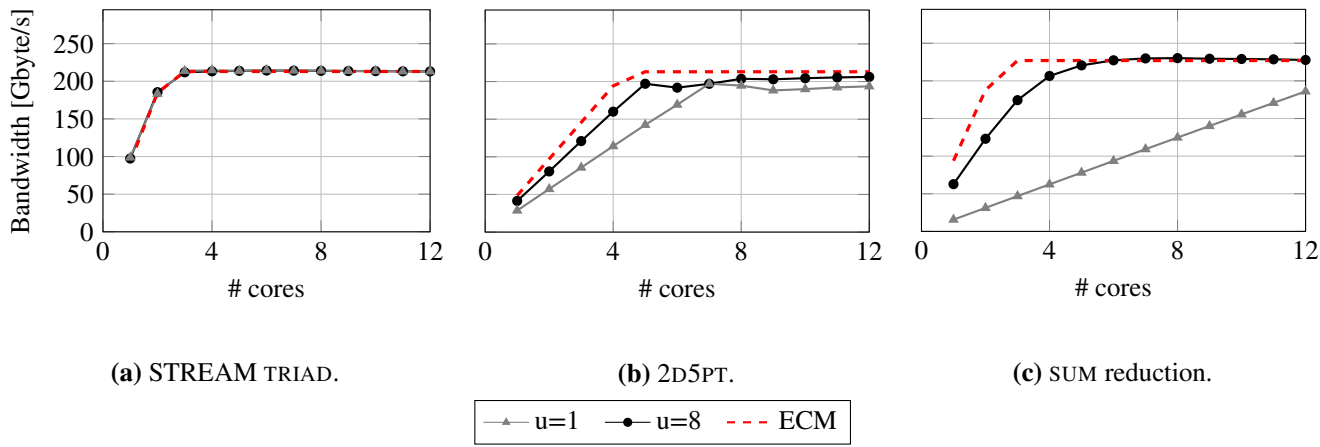

	\begin{subfigure}[tb]{0.33\textwidth}
		\centering
		%
	\input{plots/tikz/ecm/ecm_multicore_stream.tex}%

		\caption{STREAM \textsc{triad}.}
		\label{fig:ecm_mc_stream}
	\end{subfigure}
	\hspace{0.15em}
	\begin{subfigure}[tb]{0.31\textwidth}
		\centering
        \vspace{0.4em}
		%
	\input{plots/tikz/ecm/ecm_multicore_stencil.tex}%

		\caption{\textsc{2d5pt}.}
		\label{fig:ecm_mc_stencil}
	\end{subfigure}
	\hspace{0.1em}
	\begin{subfigure}[tb]{0.31\textwidth}
		\centering
        \vspace{0.35em}
		%
	\input{plots/tikz/ecm/ecm_multicore_sum.tex}%

		\caption{\textsc{sum} reduction.}
		\label{fig:ecm_mc_sum}
	\end{subfigure}
	\centering
        \smallskip
        
	%
	\input{plots/tikz/ecm/ecm_multicore_legend.tex}%

	\caption{\rAdd{Strong} scaling \rAdd{(constant working-set size)} within one \cmg\ for STREAM \textsc{triad}, \textsc{2d5pt} and \textsc{sum}
		 kernels, comparing ECM model with measurements.
		Data without unrolling are shown for reference. Note that the read-only memory bandwidth was used as a limit for \textsc{sum}.
		The working set size for \textsc{triad} and \textsc{sum} was set to $4\,\GB$.
		For \textsc{2d5pt}, the problem size was chosen as $10000^2$ so	that the layer condition is broken at L1 but fulfilled at L2.}
	\label{fig:ecm_multicore}
\end{figure}

We now move from single-core to multicore analysis within one CMG.
\Figure~\ref{fig:ecm_multicore} shows a comparison of the ECM-model predictions and measurements for the STREAM \textsc{triad}, \textsc{2d5pt} and \textsc{sum}  kernels.
While STREAM \textsc{triad} matches the prediction perfectly, for \textsc{sum} it is evident that insufficient MVE (as shown in the ``u=1'' data) is the root cause for non-saturation of the memory bandwidth due to the long ADD latency.
For the stencil kernel, saturation is possible even without unrolling, but more cores are needed.

\section{Case study: Sparse matrix-vector multiplication}\label{sec:spmvm}
Sparse matrix-vector multiplication (\spmv) is arguably one of the
most relevant numerical kernels in computational science. With the
help of the insights gained in the construction of the ECM performance
model, we are now going to analyze and optimize the performance of
\spmv\ kernels on the \afx. We restrict ourselves to general matrices
without the option of exploiting symmetries or dense substructures.

Due to its low computational intensity of at most
1/6\,\FB~\cite{Gropp:1999} (assuming double precision and four-byte
indexing), \spmv\ is \rAdd{typically} expected to be  \rAdd{memory bandwidth bound} on all modern
computer architectures if the matrix does not fit into cache.
Hence, the OpenMP-parallel kernel (i) should exhibit the typical
saturating scaling characteristics of a memory-bound code across the
cores of a \cmg, (ii) should be able to exhaust the available
memory bandwidth and (iii) should preferably show the maximum
possible computational intensity as derived in~\cite{Kreutzer14}.

In most algorithms, a left-hand-side vector $y$ is updated in the
course of the \spmv\ operation: $y=y+Ax$. Due to the lack of store
misses, zero fill instructions are thus unable to improve
the performance here. However, since there is no cache reuse
in the access to the matrix data but only in the right-hand-side
vector, the sector-cache feature may be able to restrict the cache
usage of the matrix, leaving more cache for the vector and thus
helping to get close to the maximum intensity.

\subsection{Motivation}

\begin{figure}
  \begin{minipage}[b]{0.4\textwidth}
\begin{lstlisting}
  // parallel loop
  for(int i = 0; i < N_r; ++i)
  {
    for(int j = rp[i]; j < rp[i+1]; ++j)
    {
      y[i] += val[j] * x[ci[j]];
    }
  }
\end{lstlisting}
  \captionof{lstlisting}{\crs\ \spmv\ kernel. The array \texttt{rp[]} holds the starting indices of the rows, while the array \texttt{ci[]} holds the column indices of the nonzeros. }\label{listing:crs}
\end{minipage}
\hfill
  \begin{minipage}[b]{0.55\textwidth}
\begin{lstlisting}
  // parallel loop
  for(int i = 0; i < N_r/C; ++i)
  {
    for(int j = 0; j < cl[i]; ++j)
    {
      for(int k = 0; k < C; ++k)
      {
        y[i*C+k] += val[cs[i]+j*C+k] * x[col[cs[i]+j*C+k]];
      }
    }
  }
\end{lstlisting}
  \captionof{lstlisting}{\sellcs\ \spmv\ kernel. The array \texttt{cl[]} holds the widths of the chunks, while the array \texttt{cs[]} holds their starting indices. }\label{listing:sell}
\end{minipage}
\end{figure}
\begin{figure}[tb]
\begin{subfigure}[t]{0.44\textwidth}
	\centering
	%
	\input{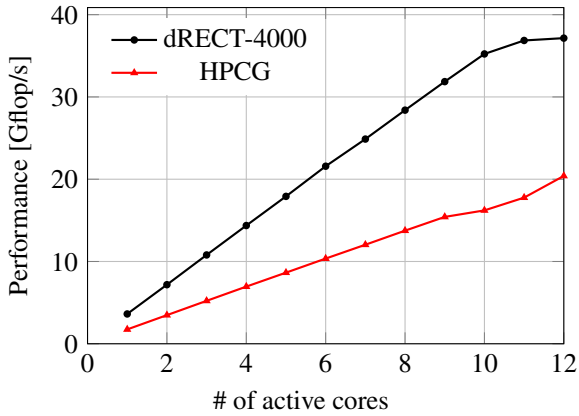}%

	\caption{\rAdd{Strong} scaling of \spmv\ with the \tands\ matrix
          ($\mbox{\NNZR}=4000$) and the HPCG matrix (problem size $128^3$)
          using the \crs\ format across the cores
          of a \cmg. }
	\label{fig:crs_gcc_scaling}
\end{subfigure}
\hspace{1em}
\begin{subfigure}[t]{0.5\textwidth}
	\centering
	%
	\input{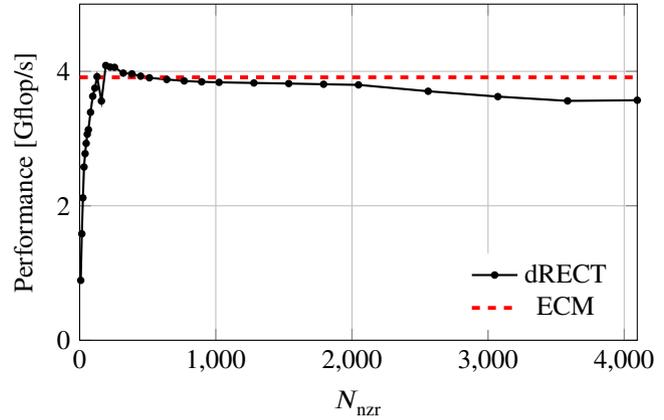}%

	\caption{Single-core performance of \spmv\ with the \tands\ matrix
          versus \NNZR. The ECM model prediction is shown for reference. The working set size for the matrix was kept constant at 500\,\MiB.}
	\label{fig:crs_gcc_size_scan}
\end{subfigure}
\caption{Performance of SpMV with CRS format, compiled using GCC.}
\end{figure}
In order to provide a baseline for experiments with realistic sparse
matrices, we start with a ``tall and skinny'' dense rectangular (\tands) matrix
stored in the Compressed Row Storage (\crs) format, also called
Compressed Sparse Row (CSR) format. \crs\ is the most popular
sparse-matrix format, and it is usually well-suited for cache-based multicore
CPUs.  Listing~\ref{listing:crs} shows the corresponding high-level loop
code. The \tands\ matrix poses no challenges in terms of load
balancing and right-hand-side access. The \rAdd{black line with symbols} in
Fig.~\ref{fig:crs_gcc_scaling} show performance scaling on one \cmg\
for a \tands\ matrix with 4000 columns,\footnote{\rAdd{For general sparse
  matrices, \NNZR\ is the average number of nonzeros per row,
  which is usually much smaller than the number of columns.
  In the special case of \tands, all rows have the same number of
  nonzeros, which equals the number of columns.}}
using GCC with plain C code. It saturates at about 37\,\GFS, which
translates to a memory bandwidth of about 220\,\GBS\ assuming the
maximum intensity of 1/6\,\FB. Hence, we observe the expected pattern,
although almost all cores are needed for saturation. In
Fig.~\ref{fig:crs_gcc_size_scan} we show a scan of the single-core
\spmv\ performance with the \tands\ matrix with respect to the number
of nonzeros per row. The sharp drop towards small \NNZR\ reflects
the inefficiency of short inner loops, which we will elaborate on
later.

Unfortunately, the \tands\ case is not representative of most
realistic sparse matrices, even for those with ``benign'' structures.
The red line in Fig.~\ref{fig:crs_gcc_scaling} shows performance
scaling for the HPCG matrix (problem size $128^3$, $\mbox{\NNZR}=27$).
In this case, the single-core performance is only about half
of that for the \tands\ matrix, thus bandwidth saturation cannot
be achieved.  The cause of this failure is the generated
assembly code: Although the compiler can vectorize the inner
kernel along the matrix row, it accumulates the results into
a single target register, which incurs the full \verb.fmla.
latency of 9\,\cycles\ in every SIMD loop iteration. At $\mbox{\NNZR}=27$,
the inner loop has four iterations. Together with the latency of
the required horizontal add instruction (\verb.faddv.) of 49\,\cycles,
one row requires $4\times 9+49\,\cycles=85\,\cycles$ to execute.
Assuming again the maximum computational intensity, this translates
into a maximum full-\cmg\ bandwidth of
\begin{equation}
  12\,\mbox{(cores)}\times 2.2\,\GCS\times 27\times 12\,\byte/85\,\cycles\approx 101\,\GBS  
\end{equation}
and a maximum performance of $16.8\,\GFS$. Note that we did not consider
the data transfers through the memory hierarchy since the overlapping
part of the in-core execution dominates strongly.
In practice, successive row executions can overlap slightly, which explains
our measurement of 20\,\GFS. Clearly the accumulation of partial sums
into a single register is part of the problem. A solution to this problem will be discussed next.

\subsection{Modulo Variable Expansion (\mve)}
\label{subsec:spmv_mve}
\begin{figure}[tb]
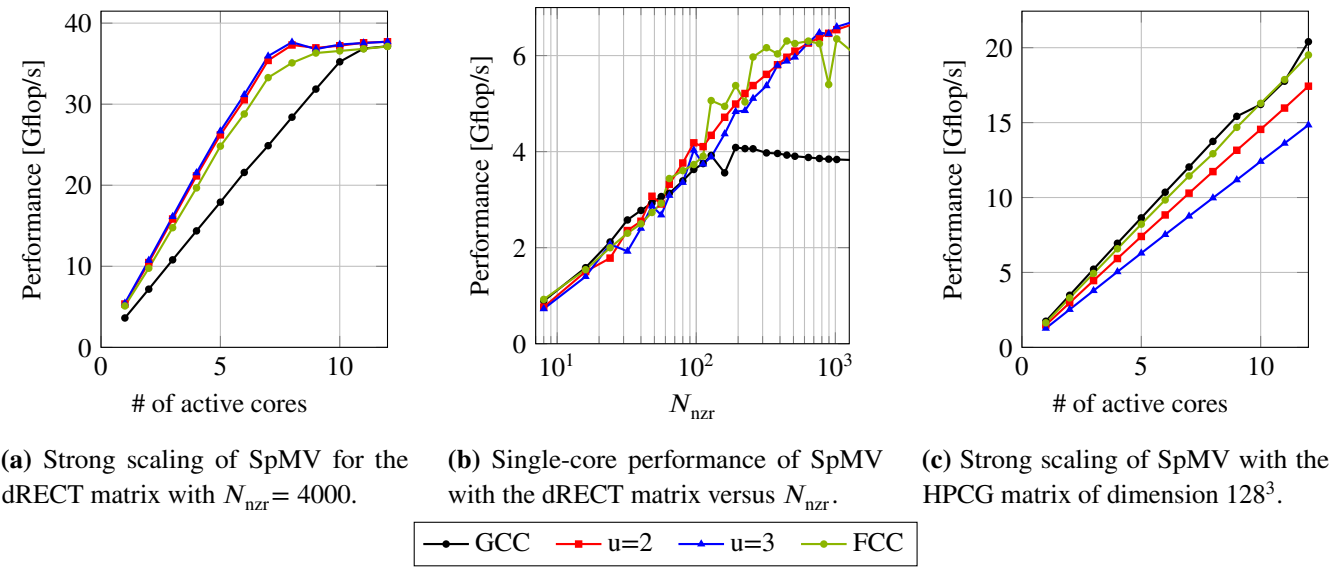

	\begin{subfigure}[t]{0.3\textwidth}
		\centering
		%
	\input{plots/tikz/spmv_microbench/gcc/scaling_mve_csr.tex}%

		\caption{\rAdd{Strong} scaling of \spmv\ for the \tands\ matrix
                  with \NNZR$=4000$.}
		\label{fig:mve_impact_scaling}
	\end{subfigure}
	\hspace{1em}
	\begin{subfigure}[t]{0.32\textwidth}
		\centering
		%
	\input{plots/tikz/spmv_microbench/gcc/single_core_nnzr_latency.tex}%

		\caption{Single-core performance of \spmv\ with the \tands\ matrix versus
                  \NNZR.}
		\label{fig:nnzr_latency_plot}
	\end{subfigure}
	\hspace{1em}
	\begin{subfigure}[t]{0.3\textwidth}
		\centering
		%
	\input{plots/tikz/spmv_microbench/gcc/scaling_mve_csr_hpcg.tex}%

		\caption{\rAdd{Strong} scaling of \spmv\ with the HPCG matrix \rAdd{of dimension $128^3$}.}
		\label{fig:mve_pwtk_scaling}
	\end{subfigure}
	\centering
	\smallskip
	
	%
	\input{plots/tikz/spmv_microbench/gcc/legend.tex}%

	\caption{Effect of \mve\ on the performance of \spmv\ using GCC with plain C code, explicit unrolling
		with GGC and \mve, and using the plain C code with the FCC compiler.}
\end{figure}

\mve~\cite{MVE} accumulates partial sums into several registers, allowing for
substantial overlapping of successive \verb.fmla. instructions. The downside
is that the computation of the final per-row result becomes more expensive since
the reduction involves more registers. 
The FCC compiler can automatically employ \mve\ and produces two code paths, with and without \mve. Which path is taken is determined at runtime depending on the inner loop length.  
The GCC compiler does not employ modulo variable expansion (\mve) even when a
\verb.#pragma unroll. directive is used. Hence, from now on  we revert to compiler
intrinsics for all unrolled kernels to exert more control over the code generation.

We start by investigating the \tands\ case.
Figure~\ref{fig:mve_impact_scaling} shows performance scaling at $\mbox{\NNZR}=4000$.
Unrolling by two or three with \mve\ clearly helps to boost the single-core
performance and thus achieves stronger saturation\footnote{Strong saturation means saturation at a number of cores much smaller than the total number of cores.} at around eight cores
with GCC\@.
As can be seen in Fig.~\ref{fig:nnzr_latency_plot}, this optimization is effective
only if \NNZR\ is not too small, because the additional overhead for the
final reduction cannot be amortized if the number of iterations
in the inner loop is small. At an intensity
of 1/6\,\FB, a single-core performance of about 3\,\GFS\ is required to
saturate the \cmg\ memory bandwidth with all cores. This becomes possible starting at
$\mbox{\NNZR}\gtrsim 50$, but a significantly higher number is necessary
to achieve strong saturation. Consequently, the \crs\ format
is unable to yield best performance for matrices from many application 
fields: Fig.~\ref{fig:mve_pwtk_scaling} shows performance scaling
with and without \mve\ for the HPCG matrix ($\mbox{\NNZR}=27$).
Saturation is not within reach.

The fundamental dilemma with the \crs\ format on \afx\ is that SIMD vectorization
and \mve\ must both be implemented within the inner loop. As a result, the inner loop becomes too short
for effective in-core latency hiding on realistic matrices. Other storage
formats such as \sellcs\ can mitigate this problem.

\subsection{\sellcs}
\label{subsec:spmv_sellcs}
\begin{figure}
  \centering
  \includegraphics*[height=0.35\textheight]{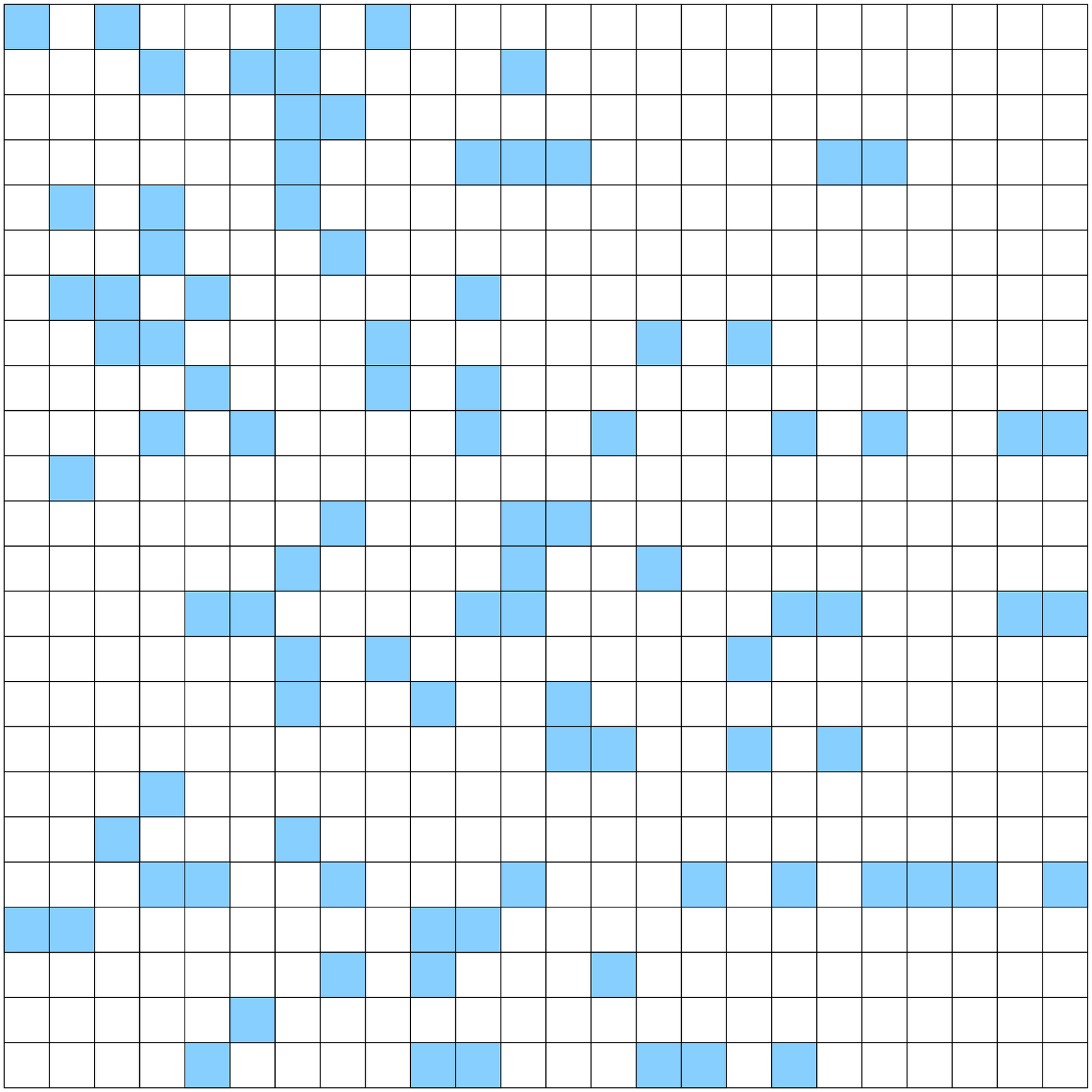}\hspace{2cm}
  \includegraphics*[height=0.35\textheight]{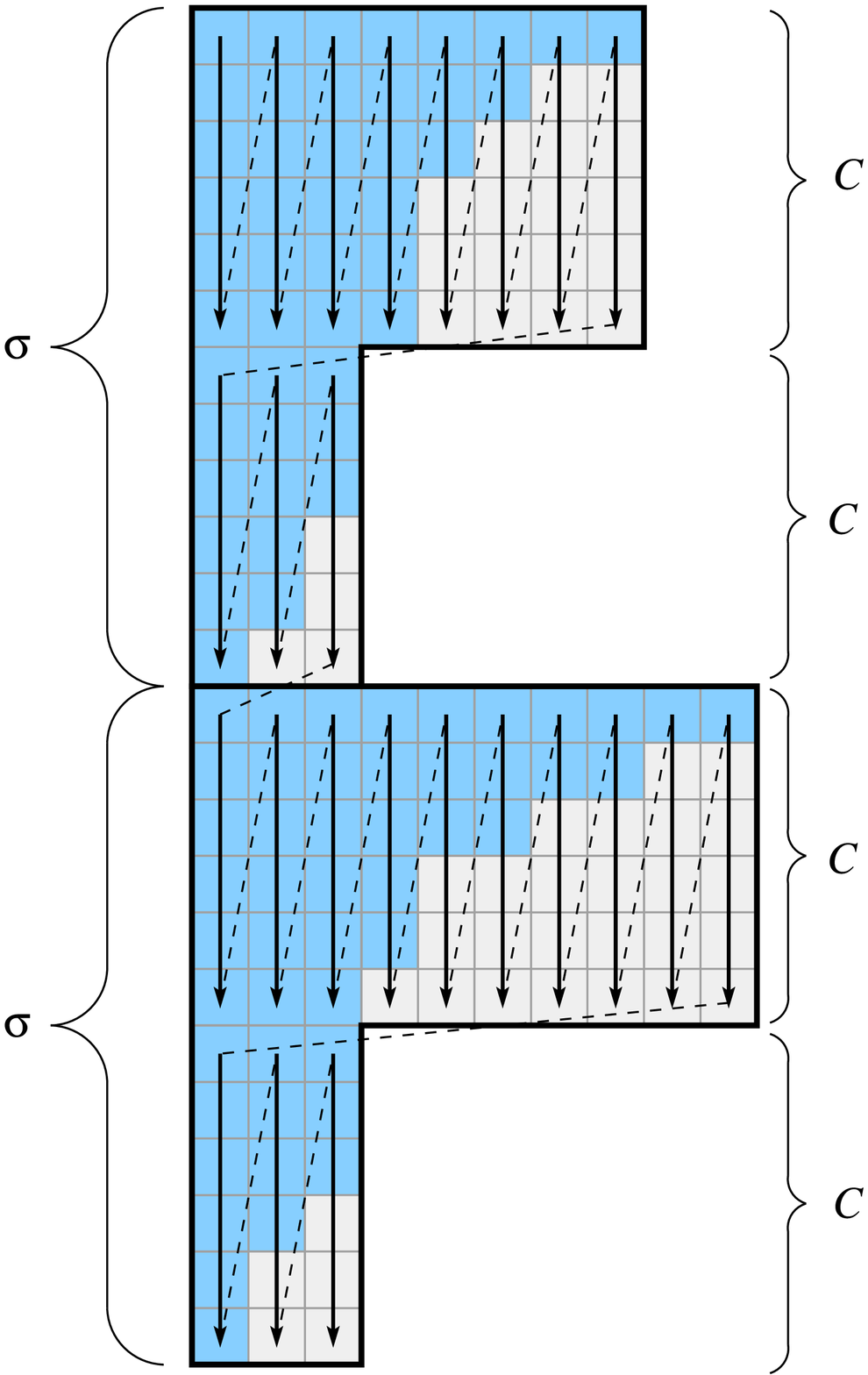}
  \captionof{figure}{A sparse matrix with $N_r=24$ rows (left) and the
    SELL-6-12 data structure generated from it (right). The blue boxes
    are nonzero entries, and the gray boxes are the zero fill-in. The
    arrows indicate the storage order. \rAdd{For illustration purposes,
      a column of nonzero entries is marked in dark blue in the SELL-6-12
      figure; the corresponding entries are shown in the original
      matrix as well. Note that the permutation is applied to row and column
      indices alike.}\label{fig:sellcs}}
\end{figure}
\sellcs~\cite{Kreutzer14} is a sparse-matrix storage format suited for
a broad range of architectures with wide SIMD or SIMT units.
To convert a matrix to \sellcs, its rows
are first sorted   within blocks of $\sigma$ rows (the \emph{sorting range}) according to the number of nonzero entries. 
Within each block of sorted rows, the nonzeros are stored in column-major
format in chunks of height $C$ (the \emph{chunk size}).
The columns of each chunk are zero-padded if necessary so that
each row within a chunk has the same length. See Fig.~\ref{fig:sellcs}
for an illustration.

The inner loop of the corresponding \spmv\ code goes over one column
of a chunk of height $C$ (see Listing~\ref{listing:sell}).
\rAdd{This means that $C$ should be a (small) multiple of the SIMD
  width, but large enough so that successive iterations of the loop,
  which accumulate into different target registers, can fill
  the bubbles in the \texttt{fmla} pipeline.}
Furthermore, no expensive horizontal reductions (\verb.faddv.) are
required. Due to the chunk padding, remainder loops cannot occur and
all SIMD lanes are filled. The second-innermost loop goes over the
columns of a chunk and is as long as the chunk width, i.e., \NNZR\ on
average. Disadvantages of the \sellcs\ format include possible
excessive zero fill-in for very irregularly-shaped matrices, and a
potential impact of the row sorting on the access to the right-hand-side vector.
\begin{figure}[tbp]
	\begin{subfigure}[t]{0.5\textwidth}
		\centering
		%
	\input{plots/tikz/spmv_microbench/gcc/single_core_nnzr_latency_sellc.tex}%

		\caption{Single-core \spmv\ performance with the \tands\ matrix
                  versus \NNZR, comparing the \crs\ format (gray, using FCC) with \sellcs\
                using different values of $C$ and compilers.}
		\label{fig:nnzr_latency_plot_sellc}
	\end{subfigure}
	\hspace{1em}
	\begin{subfigure}[t]{0.44\textwidth}
		\centering
		%
	\input{plots/tikz/spmv_microbench/gcc/scaling_sellC.tex}%

		\caption{\rAdd{Strong} scaling of \spmv\ within a \cmg\ with the HPCG matrix.}
		\label{fig:pwtk_scaling_sellc}
	\end{subfigure}
\smallskip

\centering
%
	\input{plots/tikz/spmv_microbench/gcc/legend_sellC.tex}%

	\caption{\spmv\ performance with \sellcs.}
\end{figure}

Figure~\ref{fig:nnzr_latency_plot_sellc} shows \spmv\ performance
versus \NNZR\ using the \tands\ matrix, comparing the \crs\ format (with
the FCC compiler), SELL-8-1 with GCC, and SELL-16-1 with GCC and
FCC\@. We also give the ECM-model predictions for the SELL cases.
\sellcs\ is able to keep close to the model even for small \NNZR,
owing to the advantages shown above. Even with $C=8$, which does
not allow for mitigation of the pipeline stall on the \verb.fmla.
instruction, the performance loss at low \NNZR\ is small because
of the absence of an expensive reduction after the loop across the
chunk. At $C=16$ the stall penalty is cut in half and leads to
a significant performance boost. Note also that there is a distinctive
drop in performance for the \crs\ format at $\mbox{\NNZR}\approx 4000$,
which is caused by the right-hand-side vector not fitting in the
L1 cache anymore. No such drop is visible for \sellcs\ because
the vector elements are reused along the columns of a chunk.

Figure~\ref{fig:pwtk_scaling_sellc} shows performance scaling of
\spmv\ for the HPCG matrix, comparing the same setups as in
Fig.~\ref{fig:nnzr_latency_plot_sellc}. As expected, the fastest
single-core version (SELL-16-1) also exhibits the strongest
saturation at nine cores. SELL-8-1 is also able to saturate
but requires all cores on the  \cmg.

\begin{figure}[tbp]
	\centering
	%
	\input{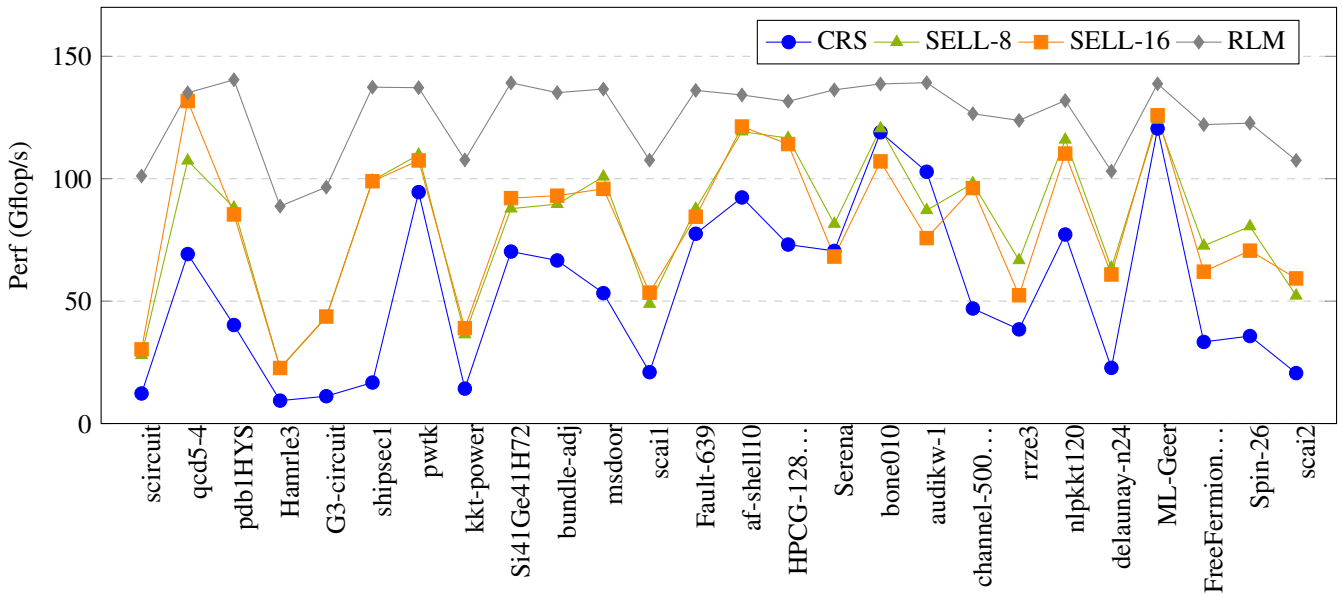}%

	\caption{Influence of the sparse-matrix storage format on the
          \spmv\ performance for CRS, SELL-8-$\sigma$, and
          SELL-16-$\sigma$ on the full \afx\ chip (48 cores).
          The best value of $\sigma$ was determined
          by exhaustive search in the range $1,\ldots,4096$
			for each matrix separately. Matrices are sorted from
         left to right in ascending order according to the number  of nonzeros.
         The sector-cache feature was not used. Diamonds show the
         matrix-specific \rl\ limits (RLM).}
	\label{fig:spmv_storage_format}
\end{figure}

Finally, we compare the \sellcs\ format with \crs\ on a range
of matrices on the full \afx\ chip (four {\cmg}s) in
Fig.~\ref{fig:spmv_storage_format}. Table~\ref{tab:test_mtx} lists
the properties of the test matrices. The matrix-specific
memory-bound \rl\ limit, i.e., assuming optimal
reuse of the right-hand-side vector~\cite{Kreutzer14},
is shown for reference. We used  three tuning parameters to find the best
per-matrix performance: (i) Reverse Cuthill-McKee~(RCM) 
reordering, (ii) row-based vs.\ nonzero-based load balancing and (iii)
 $\sigma$ in the range of 1 to 4096.

 \sellcs\ provides superior performance to \crs\
for almost all matrices. Exceptions exist where the
access pattern to the right-hand-side vector changes
for the worse compared to \crs.
The difference between $C=8$ and $C=16$ is generally small.
The significant gaps between measurement and model have
a variety of reasons: Some matrices, such as scai1 and scai2,
have a small \NNZR\ and a structure that leads to cache-unfriendly
access patterns, thereby inhibiting
saturation. In other cases, such as kkt-power, the matrix
exhibits a strong imbalance of row lengths, making load balancing
hard especially across CMGs (i.e., ccNUMA domains).

\begin{table}[tbp]
	\centering
	\caption{Details of the benchmark matrices. \NR\ is the number of
		rows, \NNZ\ is the number of nonzeros, and \NNZR\
		is the average
		number of nonzeros per row.
	Most matrices were taken from the SuiteSparse Matrix Collection~\cite{UOF}.
	The matrices with $\ast$ come from other research projects.\label{tab:test_mtx}}
	\begin{center}
		\begin{tabular}{|l|l|S[table-format=7.0, table-space-text-pre=(, table-space-text-post=)]|S[table-format=8.0, table-space-text-pre=(, table-space-text-post=)]|S[round-mode=places,round-precision=2]|}
\toprule
{Index} & {Matrix name} &  {\NR} & {\NNZ} & {\NNZR} \\
\midrule
{1} & {scircuit} &          170998 &               958936 &    5.60788 \\
{2} & {qcd5\_4} &           49152 &              1916928 &   39.00000 \\
{3} & {pdb1HYS} &           36417 &              4344765 &  119.30596 \\
{4} & {Hamrle3} &         1447360 &              5514242 &    3.80986 \\
{5} & {G3\_circuit} &         1585478 &              7660826 &    4.83187 \\
{6} & {shipsec1} &          140874 &              7813404 &   55.46378 \\
{7} & {pwtk} &          217918 &             11634424 &   53.38900 \\
{8} & {kkt\_power} &         2063494 &             14612663 &    7.08151 \\
{9} & {Si41Ge41H72} &          185639 &             15011265 &   80.86266 \\
{10} & {bundle\_adj} &          513351 &             20208051 &   39.36497 \\
{11} & {msdoor} &          415863 &             20240935 &   48.67212 \\
{12} & {scai1$\ast$} &         3405035 &             24027759 &    7.05654 \\
{13} & {Fault\_639} &          638802 &             28614564 &   44.79411 \\
{14} & {af\_shell10} &         1508065 &             52672325 &   34.92709 \\
{15} & {HPCG-128-128-128} &         2097152 &             55742968 &   26.58032 \\
{16} & {Serena} &         1391349 &             64531701 &   46.380672 \\
{17} & {bone010} &          986703 &             71666325 &   72.63211 \\
{18} & {audikw\_1} &          943695 &             77651847 &   82.28490 \\
{19} & {channel-500x100x100-b050} &         4802000 &             85362744   & 17.776498 \\
{20} & {rrze3$\ast$} &         6201600 &             92527872 &   14.92000 \\
{21} & {nlpkkt120} &         3542400 &             96845792 &   27.33903 \\
{22} & {delaunay\_n24} &       16777216 &             100663202 &  5.999994 \\
{23} & {ML\_Geer} &         1504002 &            110879972 &   73.72329 \\
{24} & {FreeFermionChain-26$\ast$} &        10400600 &            140616112 &   13.52000 \\
{25} & {Spin-26$\ast$} &        10400600 &            145608400 &   14.00000 \\
{26} & {scai2$\ast$} &        22786800 &            160222796 &    7.03139 \\
\bottomrule
\end{tabular}

	\end{center}
\end{table}

\subsection{\spmv\ and the sector cache}
\label{sec:spmv-sc}
\begin{figure}[tbp]
	\centering
	%
	\input{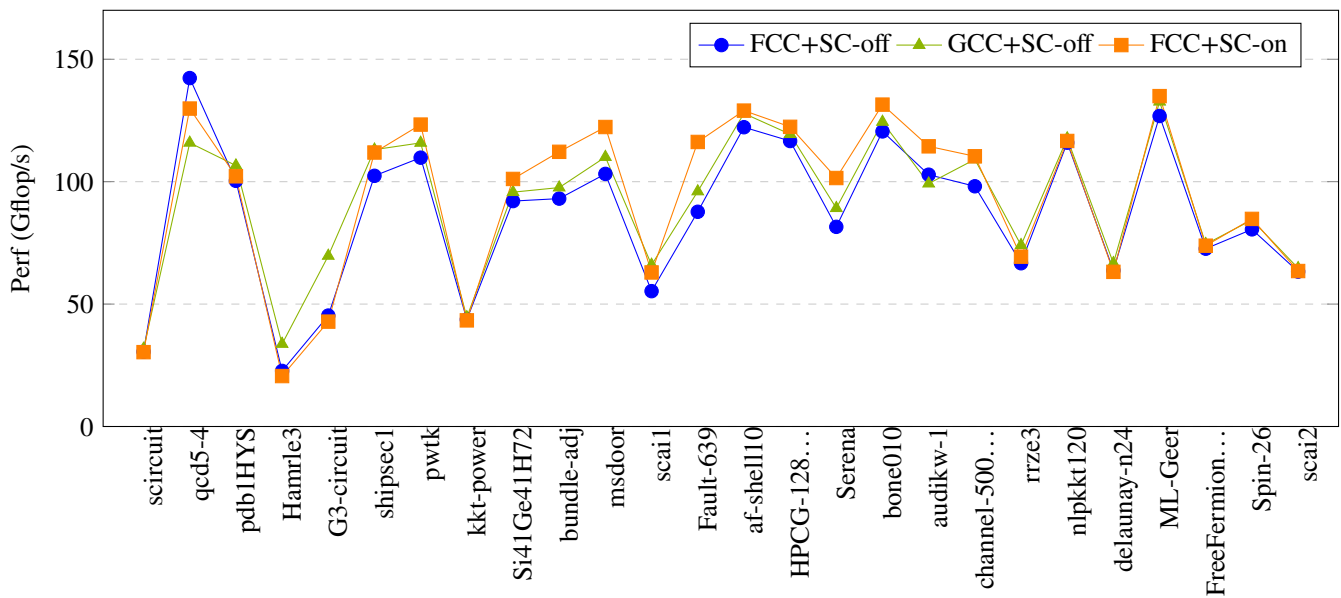}%

	\caption{Influence of sector cache on the performance of \spmv.}
	\label{fig:spmv_sc}
\end{figure}
The sector-cache feature is expected to have a positive effect on
the \spmv\ performance in cases where the cache is too small to
ensure perfect reuse of the right-hand-side vector after it is loaded
from memory. Restricting the number of cache ways used for the
matrix data leaves more space for the vector, possibly increasing the
computational intensity. Since invoking the sector cache comes
with considerable overhead (see Sec.~\ref{sec:sector-cache})
we activate it outside the repetition
loop. This is compatible with the structure of many sparse algorithms,
where the same matrix is applied repeatedly to different
vectors. Best results were obtained by allotting four ways of L2
and one way of L1 to the matrix data (nonzeros and index structures).
The tuning space  of  Sec.~\ref{subsec:spmv_sellcs} was enlarged by the chunk size ($C \in [1,128]$).

In Fig.~\ref{fig:spmv_sc} we show the impact of the sector cache
on \spmv\ performance for the test matrices, comparing with
the FCC compiler without sector cache and with GCC (which does not support sector cache). The largest
benefit is observed with medium-sized matrices, where the additional
cache space makes a difference for the right-hand-side-vector data reuse.
Matrices like qcd5-4, which just about fit in the L2, suffer because
the restricted cache space forces the matrix data into memory.

We include the GCC data in the plot because, although GCC does not
support the sector cache on the \afx, it produces better code from
the intrinsics than FCC, which can be observed for some of the smaller
matrices.

\subsection{Comparison with the Fujitsu SSL library}
\label{sec:spmv_ssl}
\begin{figure}[tbp]
	\centering
	%
	\input{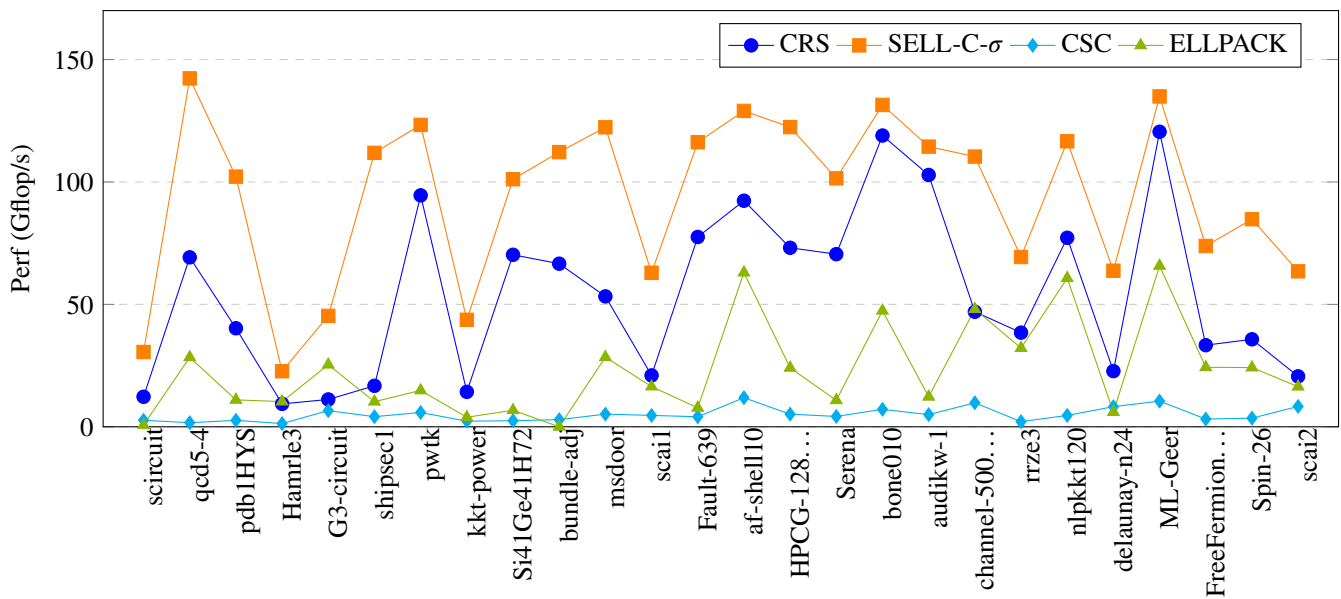}%

	\caption{Performance comparison of \spmv\ with the \rAdd{CRS format}, 
          the \sellcs\, format and the Fujitsu SSL library using
          CSC and \ellpack\ formats, respectively.}
	\label{fig:spmv_ssl}
\end{figure}

Fujitsu provides the ``C-SSL II Thread-Parallel Capabilities''
library,
which contains \spmv\ routines for a variety of formats:
Compressed Sparse Columns
(CSC, function \verb.c_dm_vmvscc.), \ellpack\ (function \verb.c_dm_vmvse.),
and Diagonal (DIA) storage. Since DIA is only suited
for matrices with diagonal structures, we ignore it here.
Figure~\ref{fig:spmv_ssl} compares our \sellcs\ \rAdd{and CRS} implementations
 with the C-SSL library. 
 The tuning space of Sec.~\ref{sec:spmv-sc} was enlarged by the sector-cache setting (on/off,  
 same number of ways as in Sec.~\ref{sec:spmv-sc}).
 The results show that
the \spmv\ implementations in 
 the current version of the C-SSL library are not competitive.

\subsection{Power consumption and tuning knobs}

\label{subsec:spmv_power}
\begin{figure}[tb]
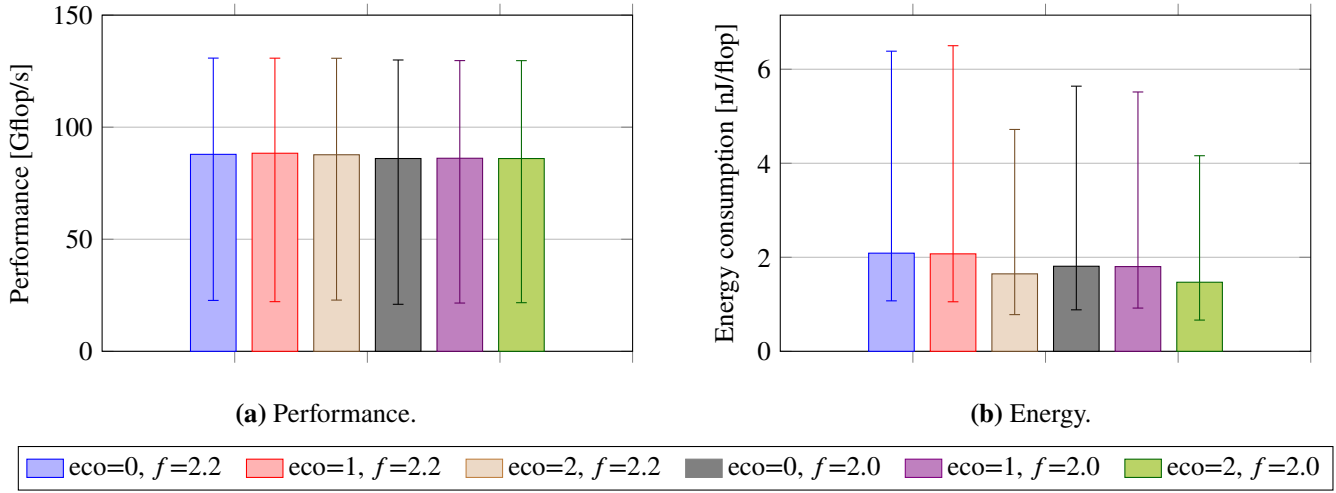

  \centering
  \begin{minipage}{\textwidth}
    \begin{subfigure}[t]{0.48\textwidth}
      %
	\input{plots/tikz/spmv_plots/power/48_threads.tex}%

      \caption{Performance.}
    \end{subfigure}
    \hfill
    \begin{subfigure}[t]{0.48\textwidth}
      %
	\input{plots/tikz/spmv_plots/power/48_threads_power.tex}%

      \caption{Energy.}
    \end{subfigure}
  \end{minipage}
  \smallskip
  
  \begin{minipage}{\textwidth}
    \centering
    %
	\input{plots/tikz/spmv_plots/power/legend.tex}%

  \end{minipage}
  \caption{Comparison of the effects of different power settings
    on node performance and energy consumption
    for \spmv, reporting the median (bars), minimum and maximum (whiskers)
    over all benchmark matrices.
    The power dissipation varies between 130\,\W\ and 190\,\W\ for
    the ``hottest'' setting ($\mathrm{eco}=0$ and $f=2.2\,\GHZ$) and
    between 76\,\W\ and 133\,\W\ for the ``coolest''
    setting ($\mathrm{eco}=2, f=2.0\,\GHZ$).}
  \label{fig:spmv_power}
\end{figure}
We explored two of the tuning knobs provided by the  FX1000 system to optimize the energy
consumption of the \afx\ processor: the clock speed ($2.0\,\GHZ$ or
$2.2\,\GHZ$) and the ``eco'' setting, which can be 0 (disabled), 1, or 2.
For a strongly memory-bound code like the \sellcs\ variant
of \spmv\ on ``benign'' matrices, we expect that lowering
the clock speed will have a 
\rAdd{negligible} 
impact on the performance
but at least a proportional influence on the power consumption
(depending on the voltage scaling, for which details are
undisclosed). Enabling eco mode, which includes disabling the FLB
floating-point unit, should also be inconsequential
for performance but probably advantageous for power.

Figure~\ref{fig:spmv_power} shows the median, maximum and minimum
performance and node energy consumption
(in \NJFLOP) for \spmv\ over all benchmark matrices,
comparing all six combinations of the three eco settings and
the two clock speeds. As expected, all settings have a minor
impact on the performance. The low-clock-speed setting reduces
the performance median by 2.7\% but lowers the energy
consumption by about 13\%.
The ``eco=2'' mode can additionally reduce the energy consumption by another 21\%,
for a total average of 31\%. In light of the fact that not all \spmv\
executions are \rAdd{memory bandwidth bound}, these are surprisingly large
savings. Although the actual energy measurements
fluctuate significantly because of the wide range of performance
numbers, the general trend is the same even for the ``hottest''
and ``coolest'' cases.

Note that all measurements were taken on a single node
of Fugaku. Significant statistical variations across nodes are expected, 
but a full coverage is beyond the scope of this paper.

\subsection{Comparison with other architectures}
\label{subsec:spmv_arch_comparison}
\begin{figure}[tb]
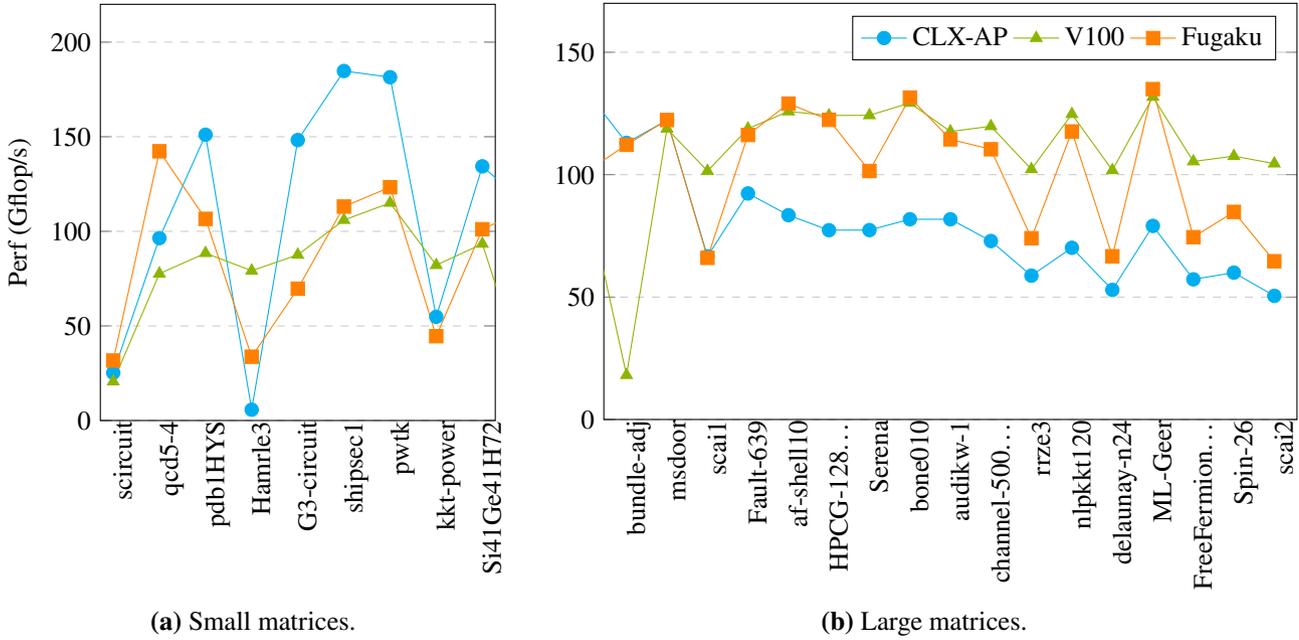

  \begin{subfigure}[t]{0.38\textwidth}
    \centering
    %
	\input{plots/tikz/spmv_plots/plot_compare_arch_cacheBound.tex}%

    \caption{Small matrices.}
    \label{fig:spmv_arch_comparsion_cachebound}
  \end{subfigure}
  \begin{subfigure}[t]{0.6\textwidth}
    \centering
    %
	\input{plots/tikz/spmv_plots/plot_compare_arch_memBound.tex}%

    \caption{Large matrices.}
    \label{fig:spmv_arch_comparsion_membound}
  \end{subfigure}
  \caption{Comparison of \spmv\ performance on \afx\ with other architectures.
    Note the different range of the vertical axis for the two graphs.}
\label{fig:spmv_arch_comparsion}
\end{figure}
We choose one contemporary, high-end GPU and CPU architecture each to provide
context for the \spmv\ performance of the \afx: an NVIDIA V100 GPU and
an Intel Cascade Lake AP  (CLX-AP) node. On the V100 we use the GHOST
library~\cite{Kreutzer17} for an efficient implementation of the
\sellcs\ format. 
The search space of Sec.~\ref{sec:spmv_ssl} was extended on \afx\ by the choice of
compilers, GCC vs.\ FCC.

Figure~\ref{fig:spmv_arch_comparsion} shows the results on all three
systems, separately for ``small'' and ``large'' matrices, the boundary
being defined by the aggregate L2/L3 cache size of the CLX-AP.
Unsurprisingly, the CLX-AP performs best for most small working sets.
On in-memory matrices, Fugaku and the V100 show an approximate
1.6$\times$ -- 2$\times$ speedup with respect to CLX-AP for ``benign'' matrices,
which is in accordance with the memory-bandwidth ratio.\footnote{Note
that a Cascade Lake SP system has only half the memory
channels of CLX-AP, so we expect the speedup to double in that case.}
With irregular matrices (e.g., Hamrle3, kkt\_power, scai1/2,
Spin-26, or FreeFermionChain-26), the GPU has a clear advantage
due to its more effective latency hiding. The bundle-adj matrix
has low performance on V100 due to GHOST only supporting
row-based load balancing.

\section{Case study: Lattice QCD Domain wall kernel}\label{sec:qcd}
\subsection{Introductory remarks}

Understanding the strong interaction, one of the four known fundamental
forces in nature, is one of the major challenges in physics.  Quantum Chromodynamics
(QCD) is the quantum field theory of the strong interaction, which describes the interaction of quarks and gluons.  
Lattice QCD is a
computer-friendly version of QCD, in which simulations are carried out on a regular
lattice in Euclidean space-time.  State-of-the-art research in Lattice QCD
requires supercomputers such as Summit or Fugaku.

A significant part of the CPU time in Lattice QCD simulations is spent on solving a
linear system of equations using iterative (multi-grid) techniques.
The key computational kernel is the application of the lattice Dirac operator to a
quark-field vector $\Psi$.  The quark field $\Psi(n)_{\alpha a}$ defined at lattice
site $n$ in a four-dimensional volume $V_4 = L_x\times L_y\times L_z\times L_t$ carries a spinor
index $\alpha = 1,2,3,4$ and a color index $a = 1,2,3$.  The interaction
is represented by SU(3) matrices $U_\mu(n)$, $n \in V_4$, carrying color indices.
This matrices are defined on the links between adjacent sites $n$ and $n+\hat{\mu}$, where $\hat{\mu}$
is the unit vector in direction $\mu$.

Multiple formulations of quarks on the lattice exist.
In this work we focus on the domain wall (DW) fermion formulation \cite{KAPLAN1992342}, in which the
physical quark field $\Psi(n)$ lives on the four-dimensional boundary of a five-dimensional
space-time lattice with
volume $V_4 \times L_s$.  The formulation involves a fermion field $\psi(n,s)_{\alpha a}$
that lives in five dimensions and carries an additional index $s = 1,\ldots,L_s$.
The interaction $U_\mu(n)$ does not depend on the fifth dimension and is replicated
along the $s$-direction.  The performance-relevant part $D$ of the domain wall Dirac operator
acts on the fermion field as follows,\footnote{The full operator is given in \cite{Furman:1994ky}, Eqs.~(2.5)--(2.7). Our $D$ corresponds to their $D^{||}$ with $M=4$.}
\begin{equation}
\psi'(n,s)_{\alpha a}= (D \psi)(n,s)_{\alpha a}=\sum\limits_{\mu=1}^4 \sum\limits_{\beta=1}^4 \sum\limits_{b=1}^3
\left\{U_\mu(n)_{ab}(1+\gamma_\mu)_{\alpha\beta}\psi(n+\hat{\mu}, s)_{\beta b} +
U^\dagger_\mu(n-\hat{\mu})_{ab} (1-\gamma_\mu)_{\alpha\beta}\psi(n-\hat{\mu},s)_{\beta b}\right\}.
\label{eq:dwf}
\end{equation}
Here, the $\gamma_\mu$ are constant $4\times 4$ Dirac matrices carrying spinor
indices.  The result of the projection $(1\pm\gamma_\mu)\psi$ is a four-component
spinor for each color.  Up to multiplicative factors only two components each are independent.
The number of input operands per site is $8\times 9$ for the $U$-fields and $8\times 12$
for the $\psi$-fields.  The number of output operands is $1 \times 12$ for
the $\psi'$-field per site. All these operands are complex numbers.

\begin{figure}
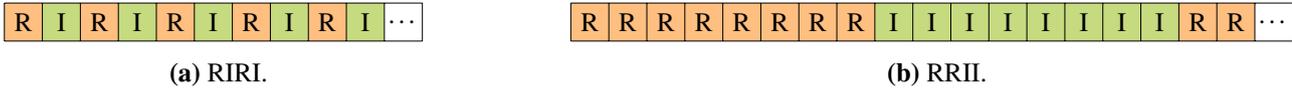

	\centering
	\begin{subfigure}[t]{0.32\textwidth}
		\centering
		%
	\input{plots/tikz/qcd/riri_illustration.tex}%

		\caption{RIRI.} \label{fig:riri_layout}
	\end{subfigure}
	\hfill
	\begin{subfigure}[t]{0.62\textwidth}
		\centering
		%
	\input{plots/tikz/qcd/rrii_illustration.tex}%
	
		\caption{RRII.} \label{fig:rrii_layout}
	\end{subfigure}
	\caption{\rAdd{Illustration of RIRI and RRII data layouts for a 512-bit SIMD width. R's refer to real and I's to imaginary parts of double-precision complex numbers. 
}}
	\label{fig:riri_vs_rrii_layout_illustration}
\end{figure}

Grid \cite{Boyle:2015tjk} is a Lattice QCD software framework written in C++
with OpenMP and MPI parallelization.  A variety of architectures are supported, including
all Intel x86 SIMD extensions, Arm NEON and 512-bit SVE~\cite{Meyer:cluster18, Meyer:2019gbz} and GPGPUs.
Grid achieves 100\% SIMD efficiency on all architectures by combining
template meta-programming and intrinsics where available.  The data layout for
complex numbers interleaves real and imaginary parts, i.e., the numbers are stored as RIRI, where R/I stands for real/imaginary part \rAdd{(see Fig.~\ref{fig:riri_layout})}.
Using the SVE instruction set, hardware processing of  $z_1 \pm z_2\cdot z_3$ (complex multiply-add)
and $z_1 \pm iz_2$ is implemented using the \texttt{svcmla} and \texttt{svcadd} ACLE intrinsics, respectively.
Key computational kernels such \rAdd{as the one emerging from} Eq.~\eqref{eq:dwf} have been specialized for the \afx~\cite{Meyer:aplat2020}.

In this work we study the performance of the DW kernel for different compilers. We also compare the interleaved
data layout with an alternative ``split'' layout, where the real and imaginary parts are stored as RRII such that the R's and I's end up in separate vector registers. For example, for double precision and a 512-bit SIMD width, we have eight consecutive R's \rAdd{as shown in Fig.~\ref{fig:rrii_layout}}. 
We use a subset of Grid~\cite{Boyle:gridbench}, which we
extended for studying CLX-AP and the \afx~\cite{Meyer:gridbench}. CLX-AP
only supports real arithmetics, therefore the interleaved layout implies permutations.
On the \afx\ we use hardware support for the computation of the interleaved layout and real arithmetics otherwise.

\subsection{Code analysis}
\label{subsec:qcd_code_analysis}

\begin{figure}[tb]
	\begin{minipage}{\textwidth}
		\begin{lstlisting}[mathescape]
		#define x_p 1 // x-plus  direction
		#define x_m 2 // x-minus direction
		#define y_p 3 // y-plus  direction
		...
		#pragma omp parallel for schedule(static)
		for {t,z,y,x} = 1:{$L_t$-2,$L_z$-2,$L_y$-2,$L_x$-2}//collapsed loop over 4d space-time
		{
			for(int s=0; s<$L_s$; ++s) //loop over 5th dimension
			{
				O[t][z][y][x][s] = R(x_p)$\,$$\cdot$$\,$U[x_p][t][z][y][x]$\,$$\cdot$$\,$P(x_p)$\,$$\cdot$$\,$I[t][z][y][x+1][s] +
                           R(x_m)$\,$$\cdot$$\,$U[x_m][t][z][y][x]$\,$$\cdot$$\,$P(x_m)$\,$$\cdot$$\,$I[t][z][y][x-1][s] +
                           R(y_p)$\,$$\cdot$$\,$U[y_p][t][z][y][x]$\,$$\cdot$$\,$P(y_p)$\,$$\cdot$$\,$I[t][z][y+1][x][s] +
                           R(y_m)$\,$$\cdot$$\,$U[y_m][t][z][y][x]$\,$$\cdot$$\,$P(y_m)$\,$$\cdot$$\,$I[t][z][y-1][x][s] +
                           R(z_p)$\,$$\cdot$$\,$U[z_p][t][z][y][x]$\,$$\cdot$$\,$P(z_p)$\,$$\cdot$$\,$I[t][z+1][y][x][s] +
                           R(z_m)$\,$$\cdot$$\,$U[z_m][t][z][y][x]$\,$$\cdot$$\,$P(z_m)$\,$$\cdot$$\,$I[t][z-1][y][x][s] +
                           R(t_p)$\,$$\cdot$$\,$U[t_p][t][z][y][x]$\,$$\cdot$$\,$P(t_p)$\,$$\cdot$$\,$I[t+1][z][y][x][s] +
                           R(t_m)$\,$$\cdot$$\,$U[t_m][t][z][y][x]$\,$$\cdot$$\,$P(t_m)$\,$$\cdot$$\,$I[t-1][z][y][x][s];
			}
		}
		\end{lstlisting}
		\captionof{lstlisting}{Simplified view of the domain wall kernel.
			$L_x$, $L_y$, $L_z$ and $L_t$ are the lattice sizes in the $x$, $y$, $z$ and $t$
			dimensions, respectively, and $s$ is the innermost fifth dimension with extent $L_s$.
		}
		\label{lst:qcd_code}
	\end{minipage}
\end{figure}

The domain wall kernel Eq.~\eqref{eq:dwf} is a radius-1 star-shaped stencil~\cite{INSPECT}
without the center element.  The input and output of the stencil operation
are the fermion fields $\psi(n,s)$ and $\psi'(n,s)$, respectively.
The interaction matrices  $U_\mu(n)$ and their inverses $U^\dagger_\mu(n)$ can
be considered as variable stencil coefficients.
Listing~\ref{lst:qcd_code} shows a simplified version of the DW kernel omitting
color and spinor indices as well as boundary conditions.
The stencil code loops over the
four dimensions $x$, $y$, $z$ and $t$, and the fifth dimension $s$.
The input fermion $\psi(n,s)$ is stored in the array \texttt{I}, and the output
fermion $\psi'(n,s)$ in the array \texttt{O}. The interaction matrices
$U$ and $U^\dagger$ are stored in \texttt{U} for the
 forward and backward directions  in $x$, $y$, $z$ and $t$.
Application of $(1\pm\gamma_\mu)$ to $\psi$ is arranged in two parts:
spinor projection \texttt{P} and spinor reconstruction \texttt{R}.\footnote{See \cite{Joo15}
for the details of projection and reconstruction.}
The operations \texttt{P} and \texttt{R} are hard-coded and do not need any operands from memory.
For each direction, the following computational sequence is applied:
\begin{enumerate}
  \item \texttt{P} projects the $(4\times3)$-component input fermion field in \texttt{I} to a $(2\times3)$-component fermion field, where $2$ means half-spinor and $3$ is the number of colors.
  \item Two matrix-vector multiplications are applied, one for each component of the half-spinor  from step 1, using the same $3\times3$ matrix in \texttt{U}.
  \item Reconstruction \texttt{R} of the $(4\times3)$-component fermion field and addition to the output fermion field in \texttt{O} (if applicable), which are combined in one step.
\end{enumerate}
The sum of all projections in step 1 contributes $96\,\flops$.
Each matrix-vector multiplication in step 2 requires $3\cdot3 = 9$ complex multiplications and $2\cdot3 = 6$ complex additions.
Each of the complex multiplications is worth six \flops, and a complex addition is worth two \flops.
Considering two matrix-vector multiplications in step 2 we have $2\cdot(9\cdot6+6\cdot2) = 132\,\flops$ per direction.
Since there are eight directions we have a total of $8\cdot132 = 1056\,\flops$.
Reconstruction and summation of intermediate results in step 3 adds another $7\cdot4\cdot3\cdot2 = 168\,\flops$.
Thus, the theoretical total flop count is $ 96 + 1056 + 168 = 1320$.
The actual \flop\ count depends on the code implementation.
However, all performance results reported in this study will
be based on $1320\,\flops$ per lattice site update (\lup).

\begin{figure}[tb]
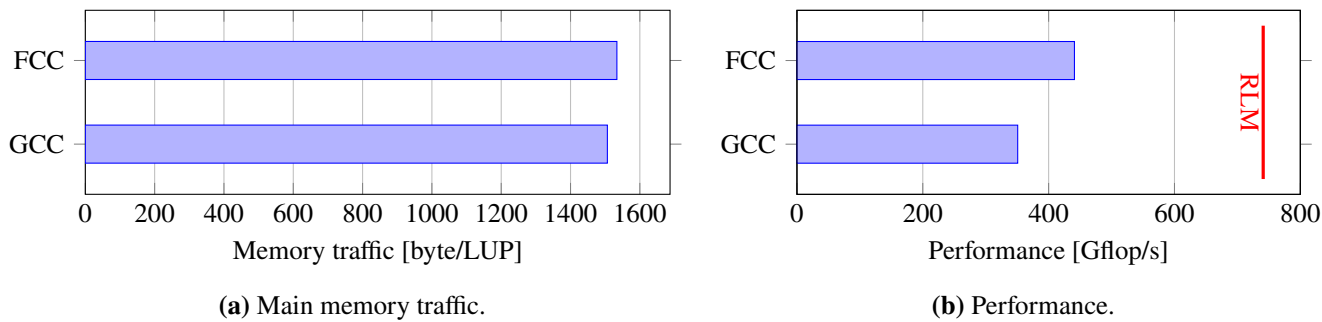

	\centering
	\begin{subfigure}[t]{0.52\textwidth}
		%
	\input{plots/tikz/qcd/base_version/48_threads_mem.tex}%

		\caption{Main memory traffic.}
		\label{fig:qcd_base_version_mem}
	\end{subfigure}
	\begin{subfigure}[t]{0.46\textwidth}
	%
	\input{plots/tikz/qcd/base_version/48_threads.tex}%

	\caption{Performance.}
	\label{fig:qcd_base_version_perf}
	\end{subfigure}
  \caption{(a) Main-memory traffic and (b) full-chip performance   of
  the baseline implementation of the DW kernel  compiled using \texttt{-Ofast} with GCC  and FCC.
	The dimension of the lattice is $24$ in each of the $x$, $y$, $z$, $t$ directions
	and $8$ in the $s$ direction.
  The \rlm\ (RLM) performance estimate is shown in (b).
}
	\label{fig:qcd_base_version}
\end{figure}

The \flop\ count along with the data traffic to and from main memory
can be used to construct a \rlm\ (RLM) performance limit.
Figure~\ref{fig:qcd_base_version_mem} shows the main-memory data traffic of a baseline implementation of the 
 DW kernel measured using the \likwidperfctr\ tool.
 This baseline implementation uses RIRI layout, ACLE intrinsics and no prefetching.
It can be seen that for the DW kernel we need approximately $1500\,\BL$.
The code intensity of the kernel can thus be estimated as $I=1320/1500\,\FB = 0.88\,\FB$.
According to the \rlm, the performance estimate is given as $\min(p_\mathrm{peak}, I\cdot b_\mathrm{Mem})$, where $p_\mathrm{peak}$ is the peak \flop\
rate and $b_\mathrm{Mem}$ is the saturated main-memory bandwidth of the hardware.
For the \afx, $p_\mathrm{peak}=3379.2\,\GFS$ and $b_\mathrm{Mem}=859\,\GBS$ (see Table~\ref{table:testbed}).
The RLM thus predicts a memory-bound performance maximum of $756\,\GFS$.

Figure~\ref{fig:qcd_base_version_perf} shows the performance of the DW
kernel on the full chip in comparison with the \rl\ prediction.
The data structure in this version of the kernel uses the interleaved (RIRI) complex array layout.\footnote{The SVE instruction set supports complex multiply-add ($z_1 + z_2\cdot z_3$), but lacks complex
multiplication ($z_1\cdot z_2$).  Therefore, each matrix-vector multiplication requires three complex multiply-add
instructions for each row, including one multiplication adding zero ($0 + z_1\cdot z_2$). The latter implies an
additional $2\cdot 3\cdot 2\cdot 8 = 96$ \flops\ on top of the 1320 \flops\ per \lup\ for the RIRI implementation.}
The code attains a performance close to $350\,\GFS$ with GCC and $440\,\GFS$ with FCC.
The measurements fall short of the RLM limit by a factor of $2.1\times$ and $1.7\times$, respectively.
In the following sections we will investigate the reasons for this
significant deviation and explore optimization strategies.

\subsection{ECM analysis, layer conditions and optimizations}
The \rlm\ predicts that the main-memory bandwidth is the performance bottleneck for the DW kernel,
but we observe almost linear scaling up to 48 cores (not shown here), indicating bottlenecks at the single-core level.
The single-core performance is $8.2$ and $10.3\,\GFS$ using GCC and FCC, respectively.
We use the ECM performance model discussed in Sec.~\ref{sec:ecm} to investigate this issue.
To construct the model, we first need
the ECM contributions \TOL, $T_\mathrm{L1\_LD}$, $T_\mathrm{L1\_ST}$, $T_\mathrm{L2}$ and $T_\mathrm{Mem}$ (see Sec.~\ref{subsec:ecm_contrib}). These must be combined using the overlap hypothesis described in Sec.~\ref{subsec:overlap_hypo}.

The L1-to-register contributions $T_\mathrm{L1\_LD}$, $T_\mathrm{L1\_ST}$ and the in-core computational contribution \TOL\
can be estimated by analyzing the assembly code with the OSACA tool. 
For the data delays in the memory hierarchy, i.e., $T_\mathrm{L2}$ and $T_\mathrm{Mem}$,
we need the data volume $V_i$ transferred over each data path $i$ and the corresponding
hardware bandwidth  $b_i$ (see Sec.~\ref{sec:ecm}) to plug into Eq.~\eqref{eq:ecm_contrib_ti}.
The data traffic can be modeled analytically,
which can be done for stencil codes using layer conditions \cite{lc_origin,sthw15,kerncraft}.
Comparison of the prediction with performance counter measurements
helps to validate the model and can reveal bottlenecks due to the code implementation and/or compiler code generation.

\subsubsection{Layer conditions}
\label{subsec:qcd_lc}
Layer conditions (LC)  provide important information about
which cache can hold which elements  of the stencil and about
the amount of data that must be transferred from and to a cache.
\rAdd{The concept is based on reuse distance analysis, i.e., 
	the distance after which a certain element of the stencil
	is reused. 
	Since stencils have a well-defined regular access pattern,
	this reuse distance can be determined analytically.}
The LC analysis assumes
 caches with infinite associativity and
least recently used (LRU) replacement policy.
Real caches, such as on the \afx, have finite associativity and might only
implement a pseudo-LRU policy.
However, it has been shown in~\cite{sthw15,yasksite} that for most stencil
codes these assumptions do not hamper the quality of the
predictions. 
For simplicity we assume in the following that the lattice 
is sufficiently large such that the working set does not fit into cache.

\paragraph{Data traffic analysis}
To construct the LC we need to analyze the data access patterns in the stencil code
(Listing~\ref{lst:qcd_code}).
We can neglect projections \texttt{P}
and reconstructions \texttt{R} since these do not contribute to
data traffic. 
The innermost loop is along the $s$ dimension, followed
by the $x$, $y$, $z$ and $t$ dimensions.
\rAdd{The elements of the array \texttt{U} are $3\times 3$ matrices,
  whose entries are of type \texttt{double} \texttt{complex}  (16\,\bytes),
  while the elements of the arrays \texttt{I} and \texttt{O}
  are $4\times 3$ \texttt{double} \texttt{complex} matrices.
Within a \lup\ and without data reuse 
we need to load eight \texttt{U} elements of dimension $3\times3$ each and
eight \texttt{I} elements of dimension $4\times3$ each,
and then store one \texttt{O} element of dimension $4\times3$.}
In total, we touch
$(8\cdot9 + 8\cdot12 + 12)\cdot 16\,\byte = 2.88\,\KB$
in each \lup.

The elements of \rAdd{\texttt{U[$\ast$][t][z][y][x]}}\footnote{Here $\ast$ refers to all eight directions.} are independent of
the $s$ loop and are touched again  after a single \lup. 
Therefore, if a cache $i$ of size $s_i$ can hold
all the elements  required to compute a single lattice site, i.e., if $s_i > 2.88$\,\KB, then these elements can be reused in the cache $i$.
In this case, the next higher\footnote{Higher means farther away from the core, e.g., L2 is higher than L1.} memory hierarchy level $j$ has to deliver eight \texttt{U} \rAdd{elements} only once
per traversal of the $s$ loop, and therefore the data traffic $V_j$ will correspond to $(8\cdot9/L_s + 8\cdot12 + w \cdot 12)\cdot 16\,\BL$.
Here, $w$ is the write-allocate factor: We have $w=2$ if write allocation applies, which is the case in our implementations because there is neither a zero fill intrinsic nor a compiler built-in function.
The condition of optimal reuse in the $s$ dimension is labeled $LC_s$ in the following.

Once all the elements in the innermost $s$ loop are traversed, the next loop is along the $x$ dimension.
Reuse of the \rAdd{element \texttt{I[t][z][y][x-1][s]}}
can happen in this loop, since it
was touched two $x$ iterations ago.
In order to satisfy this reuse condition, a cache
has to hold all the elements touched in the $s$-loop iteration \rAdd{for} two iterations of the $x$ loop.
Therefore the $LC_x$ condition reads 
$s_i > 2 \cdot L_s \cdot (8\cdot9/L_s + 8\cdot12 + 12) \cdot 16$\,\bytes.
If this condition is satisfied by cache $i$
then memory level $j$ only has to transfer
seven elements of the array \texttt{I} instead of eight, translating to $V_j = (8\cdot9/L_s + 7\cdot12 + w \cdot 12)\cdot 16\,\BL$.

The other conditions $LC_y$, $LC_z$ and $LC_t$ are constructed along the same lines.
A summary of LC along each dimension is
shown in Table~\ref{table:qcd_lc} (see the scalar code column).
\begin{table}[!tb]
	\centering
	\caption{LC for the DW kernel for scalar and vectorized code.
		To determine the data traffic from a certain cache $i$,
		its size $s_i$ (in \bytes) has to be compared with each condition starting from
                the bottom row ($LC_t$) to top row (no reuse).
		The first condition that is met by the cache $i$ determines the  layer condition it satisfies, and the next higher hierarchy level $j$ has a data transfer volume of $V_j$.
		In our case the write-allocate factor is $w=2$.
		The $d$ factor accounts for the data layout  and will be discussed in Sec.~\ref{subsec:qcd_split}.
    For the RIRI (RRII) layout we have $d=1$ ($d=2$). }
	\label{table:qcd_lc}
	\resizebox{\linewidth}{!}{
		\begin{tabular}{| l | l | l | l |}
			\toprule
			\textbf{Name} &  \textbf{$V_j$ in \BL} & \multicolumn{2}{c|}{\textbf{$s_i >$}} \\
			\cline{3-4}
			{} & {} & {Scalar code} & {512-bit vectorized code} \\
			\hline
			{no reuse} & $(8\cdot9 + 8\cdot12 + w \cdot 12)\cdot 16$ & 0 & 0 \\
			\texttt{$LC_s$}	& $(8\cdot9/L_s + 8\cdot12 + w \cdot 12)\cdot 16$ & 2880 & 11520 \\
			\texttt{$LC_x$}	& $(8\cdot9/L_s + 7\cdot12 + w \cdot 12)\cdot 16$ &  $2 \cdot L_s (8\cdot9/L_s + 8\cdot12 + 12) \cdot 16$ & $d \cdot 8 \cdot L_s (8\cdot9/L_s + 8\cdot12 + 12) \cdot 16$ \\ 
			\texttt{$LC_y$}	& $(8\cdot9/L_s + 5\cdot12 + w \cdot 12)\cdot 16$ &  $2 \cdot L_s \cdot L_x (8\cdot9/L_s + 7\cdot12 + 12) \cdot 16$ & $d \cdot 8 \cdot L_s \cdot L_x (8\cdot9/L_s + 7\cdot12 + 12) \cdot 16$ \\ 
			\texttt{$LC_z$}	& $(8\cdot9/L_s + 3\cdot12 + w \cdot 12)\cdot 16$ &  $2 \cdot L_s \cdot L_x \cdot L_y (8\cdot9/L_s + 5\cdot12 + 12) \cdot 16$ & $8 \cdot L_s \cdot L_x \cdot L_y (8\cdot9/L_s + 5\cdot12 + 12) \cdot 16$ \\ 
			\texttt{$LC_t$}	& $(8\cdot9/L_s + 1 \cdot 12 + w \cdot 12)\cdot 16$ &  $2 \cdot L_s \cdot L_x \cdot L_y \cdot L_z (8\cdot9/L_s + 3\cdot12 + 12) \cdot 16$ & $4 \cdot L_s \cdot L_x \cdot L_y \cdot L_z (8\cdot9/L_s + 3\cdot12 + 12) \cdot 16$ \\ 
			\bottomrule
		\end{tabular}
	}
\end{table}

\begin{figure}[tb]
	\begin{minipage}{0.46\textwidth}
		\centering
		%
	\input{plots/tikz/qcd//vectorization_illustration.tex}%

		\caption{Illustration of the vectorization scheme for a lattice of size $8$ each in the outermost dimensions $t$ and $z$. Complex elements from four different
			partitions are packed into one SIMD register.
			The  ordering of lattice sites is shown with numbers.
		}
		\label{fig:qcd_vectorization}
	\end{minipage}
	\hspace{1em}
	\begin{minipage}{0.52\textwidth}
		\centering
        \vspace{0.5em}
		%
	\input{plots/tikz/qcd//data_sharing_btw_cores.tex}%

		\caption{Illustration of data sharing between four cores
			attached to the same shared cache for a local (virtual) lattice size of
			$8$ each in the outermost dimensions $t$ and $z$ with periodic boundary
                        conditions. Each color represents
			a different core.  The dark-colored elements show
			the  stencil accesses at a specific time.
			We observe that two cores access the same element in the $t$ direction.
		}
		\label{fig:qcd_data_sharing}
	\end{minipage}
\end{figure}

\paragraph{Serial code and vectorization}

Stencil codes are typically vectorized along the innermost dimension.
However, in the Grid Lattice QCD framework~\cite{Boyle:2015tjk}, vectorization is implemented along the
4d space-time dimensions for two reasons: First, the extent $L_s$ of the fifth dimension would have to be a multiple of the vector length (VL=4 for the RIRI layout in double precision), which
imposes restrictions on the choice of $L_s$ that are specific to the underlying hardware architecture.
Second, other Lattice QCD kernels do not use a fifth dimension and use the vectorization scheme in 4d  space-time dimensions.

 An MPI-style 
partitioning scheme is applied to define the mapping of lattice sites to
SIMD registers.
One 512-bit SIMD register of the \afx\ can hold four complex numbers in double precision; therefore,
the lattice is divided into four partitions.
This is realized by cutting the outermost two dimensions in the 4d
space-time in half, i.e., each of the four local (virtual) partitions has a
size of $L_x \times L_y \times L_z/2 \times L_t/2 \times L_s$.
Figure~\ref{fig:qcd_vectorization} illustrates the partitioning of the lattice
and the mapping of lattice sites to SIMD registers.
Note that the site numbering is not lexicographic: Adjacent elements
of a SIMD register belong to different partitions.
Therefore, a change in site numbering implies a change in the data access pattern, which must be taken into account in the construction of the LC.
Access to all four
partitions is identical, and we can account for this by only considering the access pattern within
one local partition and multiplying the number of elements touched
at each access by a factor of four. 
For instance, the $LC_t$ condition in Table~\ref{table:qcd_lc} becomes
$s_i > \textbf{4} \cdot 2 \cdot L_x \cdot L_y \cdot  (L_z/\textbf{2}) \cdot L_s \cdot (8\cdot9/L_s + 3\cdot12 + 12) \cdot 16$\,\bytes.
The highlighted factors $4$ and $2$ reflect
the changes due to vectorization.
LC for scalar and vectorized
code are shown in Table~\ref{table:qcd_lc}.
It can be seen that this kind of vectorization
makes the layer conditions more stringent. However,
compared to the traditional inner-loop vectorization approach it
does not imply redundant L1-to-register
loads and/or shuffle operations.

\begin{figure}[tb]
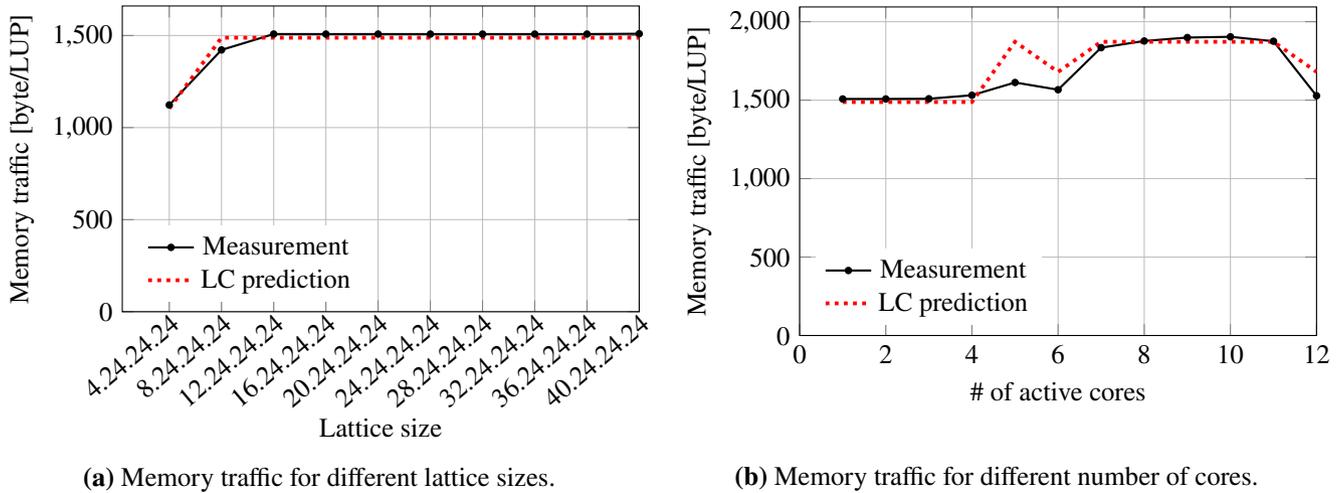

	\begin{subfigure}[tb]{0.47\textwidth}
		\centering
        \vspace{0.15em}
		%
	\input{plots/tikz/qcd/final_opt/size_scan_mem_1_thread.tex}%

		\caption{Memory traffic for different lattice sizes.}
		\label{fig:lc_size_riri}
	\end{subfigure}
	\hspace{1em}
	\begin{subfigure}[tb]{0.47\textwidth}
	\centering
	%
	\input{plots/tikz/qcd/final_opt/scaling_mem_riri.tex}%

	\caption{Memory traffic for different number of cores.}
	\label{fig:lc_threads_riri}
	\end{subfigure}
	\caption{(a) Influence of lattice size on memory traffic on a single core with $L_s=8$ and (b) influence of number of cores on memory traffic for lattice size $24^4\times8$.
  Predictions are shown as red dashed lines.
  }
	\label{fig:lc_riri}
\end{figure}

Figure~\ref{fig:lc_size_riri} shows the LC effect by
changing the lattice size $L_x$ on a single core, keeping the extent of the other dimensions constant.
For $L_x = 4$ the L2 cache satisfies $LC_z$.
As $L_x$ increases, the L2 cache violates the $LC_z$ condition (see the linear dependence
between $LC_z$ and $L_x$ in Table~\ref{table:qcd_lc}).

\paragraph{Multicore layer conditions}
The basic LC principle applied to serial code  also
holds for multicore. However, care should be taken when
the cache $i$ under consideration is a shared cache.
In this case, the per-core cache size $s_i$ and possible
sharing of data between cores has to be taken into account.
The first aspect is easily integrated into LC
by dividing the size $S_i$ of the shared cache by the number of active cores $n$, i.e., $s_i = S_i/n$.
Proper handling of data sharing among cores requires knowledge of
the loop-scheduling technique and the lattice dimensions.
For the DW kernel, thread parallelization is done via
default OpenMP static scheduling along the outermost loop
over the collapsed 4d space-time dimensions (see Listing~\ref{lst:qcd_code}).
Thus in most cases there is no data sharing among
the cores. However, the data traffic can be reduced by sharing of elements of the stencil array \texttt{I}.
This is the case, e.g., when the lattice dimensions are chosen to be a small multiple of the number of cores that share the same cache.
 Figure~\ref{fig:qcd_data_sharing} illustrates data sharing between cores. 
 In the outermost direction $t$ two cores share a stencil element.
These sharing effects have to be taken into account when
 adapting the LC to the multicore environment.

Figure~\ref{fig:lc_threads_riri} shows the influence
of the number of cores on the LC due to the shared L2 cache.
The lattice size is $24^4\times8$, i.e., a local (virtual) partition has a size of  $24\times24\times12\times12$ in the 4d space-time dimensions.
Using only a single core, the L2 satisfies $LC_y$.
As the number of cores increases, the available cache per
core decreases. At five cores, $LC_y$ is violated for L2, and only
$LC_x$ is satisfied. This explains the increase in the traffic between main memory and L2.
At six cores, the traffic decreases due to sharing of the data between cores as the local outermost dimension (12) is a small multiple of six and a condition similar to
Fig.~\ref{fig:qcd_data_sharing} applies.
For seven to eleven cores there is no more sharing and the traffic stays at the level corresponding to
$LC_x$. At twelve cores it decreases again
due to sharing.
Note that the model (shown with dotted lines) suffers some loss in accuracy especially at the boundaries of LC transitions.
This is due to the assumptions made in the model, i.e., infinite cache associativity with LRU and perfectly synchronized ``lockstep'' execution across cores.

\subsubsection{Initial optimizations}
Before diving into ECM performance predictions we first look into
initial optimizations based on the LC and
the insights gained from the microarchitecture analysis (see Sec.~\ref{ssec:incore}).

\begin{figure}[tb]
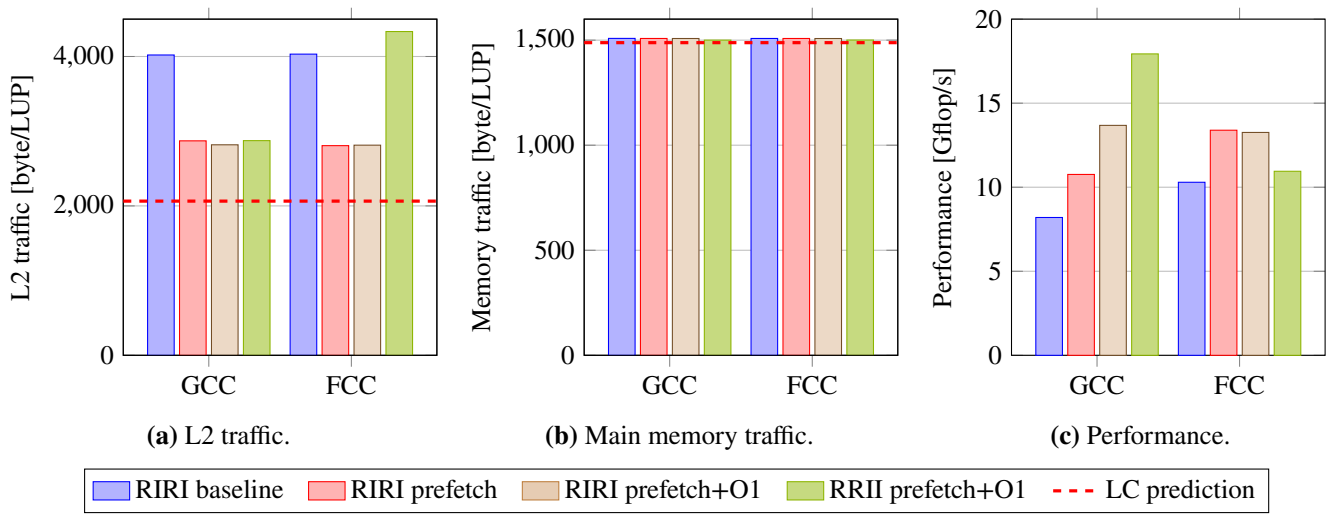

	\centering
	\begin{minipage}{\textwidth}
		\begin{subfigure}[t]{0.32\textwidth}
			%
	\input{plots/tikz/qcd/prefetch_reorder_version/1_thread_l2.tex}%

			\caption{L2 traffic.}
			\label{fig:qcd_prefetch_reorder_version_l2_scalar}
		\end{subfigure}
		\hfill
		\begin{subfigure}[t]{0.32\textwidth}
			%
	\input{plots/tikz/qcd/prefetch_reorder_version/1_thread_mem.tex}%

			\caption{Main memory traffic.}
			\label{fig:qcd_prefetch_reorder_version_mem_scalar}
		\end{subfigure}
		\hfill
		\begin{subfigure}[t]{0.32\textwidth}
			%
	\input{plots/tikz/qcd/prefetch_reorder_version/1_thread.tex}%

			\caption{Performance.}
			\label{fig:qcd_prefetch_reorder_version_perf_scalar}
			\end{subfigure}
	\end{minipage}
	\smallskip

	\begin{minipage}{\textwidth}
		\centering
		%
	\input{plots/tikz/qcd/prefetch_reorder_version/legend.tex}%

	\end{minipage}
	\caption{Influence of software prefetching and compiler code generation on the DW kernel.
           Data traffic
          and predictions (red dashed lines) as well as single-core performance are shown for a lattice size of $24^4\times8$.
		The RIRI prefetch
		implementation applies software prefetching on top of the RIRI baseline implementation.
		Codes were compiled with \texttt{-Ofast}.
		The RIRI prefetch+O1 and RRII prefetch+O1 implementations were compiled with \texttt{-O1} each. 
		}
	\label{fig:qcd_prefetch_reorder_version_scalar}
\end{figure}

\paragraph{Prefetching}
Figures~\ref{fig:qcd_prefetch_reorder_version_l2_scalar} and \ref{fig:qcd_prefetch_reorder_version_mem_scalar} compare the
measured L2 and main memory traffic of the vectorized serial code with the LC predictions (shown as dashed lines).
For the baseline implementation the memory traffic is
in line with the prediction. However, the measured L2 traffic
exceeds the prediction by about $2\times$.
A closer inspection reveals that the increase in traffic
is due to the hardware prefetchers not moving the correct elements into the cache.
The DW kernel has a complex access pattern, 
which repeats only after  all the elements in the innermost loop were accessed,
i.e., after the computation of one site. This requires access to
$2.88$\,\KB\ of data for scalar code and about $11.5$\,\KB\ for 
vectorized code.
It is extremely difficult if
not impossible for the prefetchers to correctly predict the next elements.

The deficiencies of the hardware prefetching mechanics can be overcome by software prefetching.
This decreases the L2 traffic, which is now  within  30\% of the prediction
(see Fig.~\ref{fig:qcd_prefetch_reorder_version_l2_scalar}).
The single-core performance improves
by a factor of $1.3\times$ for both GCC and FCC as seen in Fig.~\ref{fig:qcd_prefetch_reorder_version_perf_scalar}.
For the following discussions we assume that software
prefetching is applied.

\paragraph{Instruction order}
In Sec.~\ref{ssec:incore} we showed that the out-of-order window of the \afx\
is small and suffers from inefficiencies in instruction reordering.
Therefore the compiler has to support the hardware in hiding instruction latencies by proper
instruction ordering.
For the DW kernel implementation guidance is given to the compiler at the level of the source code (by explicit reuse of variables and ordering of operations). 
However, we noticed that GCC rearranges the intended order of instructions
in a manner that deteriorates performance and also introduces unnecessary register spills.
In order to mitigate these undesired effects we apply 
optimization flag \texttt{-O1} instead of \texttt{-Ofast}.
This keeps the
intended order of instructions intact and also minimizes spilling.
The performance improves by $1.3\times$
as can be seen in Fig.~\ref{fig:qcd_prefetch_reorder_version_perf_scalar}.
FCC \texttt{-O1} and \texttt{-Ofast} arrange the instructions as intended and
performance is comparable to GCC \texttt{-O1}.

\subsubsection{ECM-model prediction for interleaved RIRI layout}
\label{subsec:qcd_ecm_pred}
Using the LC we can estimate the data traffic $V_i$ and derive $T_i$ using Eq.~\eqref{eq:ecm_contrib_ti}.
Along with the in-core contributions \TOL, $T_\mathrm{L1\_LD}$ and $T_\mathrm{L1\_ST}$  obtained using the OSACA tool, we can  assemble the
contributions using the overlap hypothesis (see Sec.~\ref{subsec:overlap_hypo}) to finally determine
the expected performance.
\begin{table}[!tb]
	\centering
	\caption{ECM contributions in \cycle/\lup\ to the DW kernel implementations for
          lattice size $24^4 \times 8$.
          We used the \texttt{-O1} compiler flag.
}
	\label{table:qcd_ecm_contrib}
	\begin{tabular}{l ? l | l | l | l | l | l ? l | l | l | l | l | l}
		\toprule
		{} & \multicolumn{6}{c ?}{GCC} & \multicolumn{6}{c}{FCC} \\
		\cline{2-13}
		\textbf{Implementation} & \textbf{\TOL}  & \textbf{$T_\mathrm{L1\_LD}$}  & \textbf{$T_\mathrm{L1\_ST}$}  & \textbf{$T_\mathrm{L2}$} & \textbf{$T_\mathrm{Mem}$} &  \textbf{$T_\mathrm{ECM}$}  &  \textbf{\TOL}  & \textbf{$T_\mathrm{L1\_LD}$} & \textbf{$T_\mathrm{L1\_ST}$} & \textbf{$T_\mathrm{L2}$} & \textbf{$T_\mathrm{Mem}$} & \textbf{$T_\mathrm{ECM}$}  \\
		\hline
		\texttt{RIRI-prefetch} 		& \textcolor{black}{168.0} & 25.6 & 3 & 35.3 & 15.9 & 168.0 & \textcolor{black}{168.0} & 33 & 3 & 35.3 & 15.9 & 168.0\\
		\texttt{RRII-prefetch} 	& 70.8 & \textcolor{black}{34.4} & 20.4 & \textcolor{black}{35.3} & 15.9 & 70.8 &  85.5 & \textcolor{black}{45.2} & 37.3 & \textcolor{black}{35.3} & 15.9 & 85.5 \\
		\bottomrule
	\end{tabular}
\end{table}

\begin{figure}[tb]
	\begin{subfigure}[t]{0.48\textwidth}
		\centering
		%
	\input{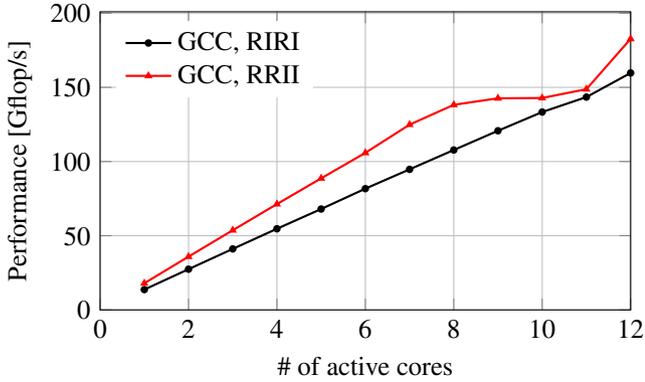}%

		\caption{\rAdd{Strong} scaling performance.}
		\label{fig:lc_threads_perf}
	\end{subfigure}
\hfill
	\begin{subfigure}[t]{0.48\textwidth}
	\centering
	%
	\input{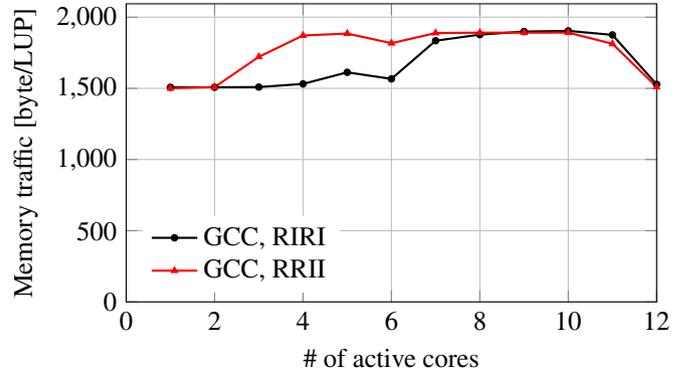}%

	\caption{Memory traffic.}
	\label{fig:lc_threads_mem}
	\end{subfigure}
	\caption{Scaling performance and main-memory traffic  of the RIRI and RRII implementations as a function of the number of active cores (GCC option \texttt{-O1}). The lattice size is $24^4\times8$.}
	\label{fig:lc_threads}
\end{figure}

For a lattice size of $24^4\times8$ we can see from Table~\ref{table:qcd_lc} that the \afx\
satisfies $LC_s$ in the L1 cache, which means  $V_\mathrm{L2} = 2064$\,\BL,
out of which $1872$\,\BL\ is due to loading data from L2 to
L1 and $192$\,\BL\ is due to storing data from L1 to L2.
The L2 load throughput is $64$\,\BC\ and the store throughput is
$32$\,\BC\ (see Table~\ref{table:testbed}), which results in
$T_\mathrm{L2} = (1872/64 + 192/32)\,\cycle/\lup = 35.25$\,\cycle/\lup.
The $T_\mathrm{Mem}$ contribution can be derived in a similar fashion.
All contributions 
are summarized in Table~\ref{table:qcd_ecm_contrib}. 
They can be combined with the overlap hypothesis (d) in Fig.~\ref{fig:triad_ovl}
to finally arrive at the ECM prediction $T_\mathrm{ECM}$. Irrespective of the compiler we find
\bq
T_\mathrm{ECM} = \max\big(168, f(25.6, 3, 35.3, 15.9)\big)\,\cycle/\lup\ = 168\,\cycle/\lup.
\eq
The $T_\mathrm{ECM}$ runtime corresponds to a single-core performance of $17.3$~\GFS.
However, we see from Fig.~\ref{fig:qcd_prefetch_reorder_version_perf_scalar}
that the measured single-core performance falls short of the prediction by 20\% and only attains about
$13.6$\,\GFS.
The deviation from the ECM model can be expected on the \afx. The
model assumes that the out-of-order execution 
overlaps multiple loop iterations and hides all latencies.
However, as the DW kernel takes almost $170$\,\cycles\ per \lup\ it is not possible for the out-of-order logic
to sufficiently overlap the iterations (see Sec.~\ref{ssec:incore}).
On a single CMG we would thus attain a performance of $12 \cdot 13.6 = 163.2$\,\GFS, which is the measured performance shown
in Fig.~\ref{fig:lc_threads_perf}.
The corresponding throughput attained by the code on a single CMG is $163.2\cdot V_\mathrm{Mem} / 1320 = 184$\,\GBS,
which does not saturate the CMG memory bandwidth
of $227$\,\GBS\ (see Table~\ref{table:testbed}).

In order to achieve saturation we have to improve the
single-core performance by at least $15$\%, which
warrants a closer look at the current bottleneck.
As can be seen in Table~\ref{table:qcd_ecm_contrib}, the bottleneck
of the RIRI implementation is the in-core overlapping part \TOL.
OSACA reveals that this 
is predominantly due to high occupation of floating-point ports on the \afx\ caused
by the high cost of SVE instructions for complex arithmetics (\texttt{fcmla} and \texttt{fcadd}). 
These instructions block the floating-point ports for three and two cycles, respectively. 
Furthermore, the \texttt{fcmla} instruction
is imbalanced between the FLA and FLB ports; two out of the three cycles are scheduled to the FLA port.
OSACA indicates that FLB has a 35\% lower occupancy than FLA.
One option to mitigate the pipeline imbalance is to avoid the SVE complex instructions and to use ordinary floating-point instructions instead. 
In this case the cost would only be two cycles for a complex
FMA and one cycle for a complex ADD operation. 
We discuss the implementation of this option in the next subsection.

\subsubsection{Split RRII layout}
\label{subsec:qcd_split}
Balanced pipeline usage is achieved by a change of the data layout.
Instead of using the interleaved RIRI layout, we switch to the split RRII layout.
In the case of 512-bit wide registers and in double precision we store
eight real parts in memory, followed by eight imaginary parts (see Fig.~\ref{fig:rrii_layout}).
Vectorized code operates on eight sites in parallel instead of four sites as in the RIRI layout (see Fig.~\ref{fig:qcd_vectorization}).
This has to be factored into the LC analysis.
The necessary modifications to the layer conditions are shown  in  Table~\ref{table:qcd_lc}, where we now have $d=2$ ($d=1$ for the RIRI layout).
This leads to tighter conditions  
and LC breaking earlier
when scaling up the number of cores. 
This is shown in Fig.~\ref{fig:lc_threads_mem}, where $LC_y$ breaks earlier with RRII than with RIRI.

The implementation of the RIRI layout guides the compiler to efficiently use the 32 floating-point vector registers without  spilling.
However, for the RRII layout the number of registers is insufficient and spilling of intermediate results  occurs.
This is reflected in the L1-to-register
contributions $T_\mathrm{L1\_LD}$ and $T_\mathrm{L1\_ST}$ shown in Table~\ref{table:qcd_ecm_contrib}.
GCC minimizes register spills when using \texttt{-O1}. FCC
produces almost $2\times$ more spills than GCC, resulting in lower performance
(see Fig.~\ref{fig:qcd_prefetch_reorder_version_perf_scalar}).
We therefore exclude FCC from further analysis.

The RRII implementation still outperforms RIRI despite the register spills.
This is because the overlapping in-core time \TOL, which is the bottleneck of the RIRI layout,
reduces by more than a factor of two (see GCC in Table~\ref{table:qcd_ecm_contrib}). 
The lower cost of ordinary (non-complex) floating-point instructions 
and the balanced pipeline usage are the main reasons for this reduction.
Still, the single-core performance falls short of the model prediction by at least a factor of 2.
We speculate that this gap is caused partly by in-core inefficiencies rooted in the dependencies between loads and stores \cite[Sec. 7.5]{a64fx_manual}, \rAdd{which are not taken into account in the  model}.
\rAdd{This can be corroborated by the 
	fact that for GCC with fewer spills the model deviates by a factor of 2.3$\times$, while for FCC the deviation is 3.1$\times$.
	Beyond the dependencies between loads and stores,
 the  inefficient OoO execution also contributes to this deviation.
To test the actual limit of the in-core execution for the code
 we constructed a  benchmark with the same instruction dependency chain as the GCC code 
and measured the runtime while keeping all the data in the L1 cache. 
This yielded 113\,\cycle/\lup, which is 1.6$\times$ higher than the in-core prediction seen  in Table~\ref{table:qcd_ecm_contrib}.
}

Using GCC, the RRII implementation already saturates the main-memory bandwidth using eight cores (see Fig.~\ref{fig:lc_threads_perf}).
The \rAdd{sudden} increase in performance \rAdd{when using} $12$ cores is due to sharing of the data in L2 among cores (discussed in Sec.~\ref{subsec:qcd_lc}).
It correlates with a drop in main-memory traffic as seen in Fig.~\ref{fig:lc_threads_mem}.
We also observe a decrease in main-memory traffic for the RIRI implementation, but it is not accompanied by an increase in performance because this version cannot saturate the memory bandwidth.

On one \cmg\ the RRII implementation achieves $182$\,\GFS, which
corresponds to a memory bandwidth of almost $205\,\GBS$.
On the full chip (four {\cmg}s) we attain $712$\,\GFS, which is a 12\% improvement over RIRI.

\subsection{Energy consumption and tuning knobs}
\begin{figure}[tb]
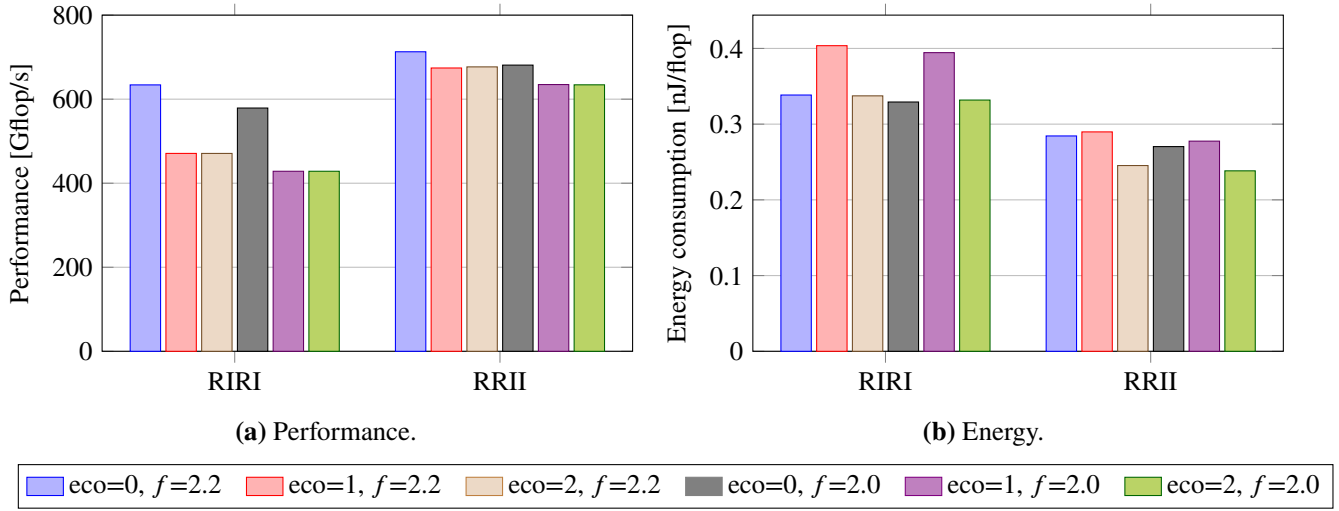

	\centering
	\begin{minipage}{\textwidth}
		\begin{subfigure}[t]{0.48\textwidth}
			%
	\input{plots/tikz/qcd/power/48_threads.tex}%

			\caption{Performance.}
			\label{fig:qcd_power_perf}
		\end{subfigure}
		\begin{subfigure}[t]{0.48\textwidth}
			%
	\input{plots/tikz/qcd/power/48_threads_power.tex}%

			\caption{Energy.}
			\label{fig:qcd_power_energy}
		\end{subfigure}
              \end{minipage}
              \smallskip
              
	\begin{minipage}{\textwidth}
		\centering
		%
	\input{plots/tikz/qcd/power/legend.tex}%

              \end{minipage}
	\caption{Comparison of performance and energy consumption
		of DW kernel implementations for different power and frequency settings on a full chip of Fugaku.
    The lattice size is $24^4\times8$. The energy consumption of the RIRI implementation is almost
    20\% higher than RRII.  The energy consumption of RRII can be reduced without significant loss in performance
    by lowering the clock frequency and switching on eco mode.
    }
	\label{fig:qcd_power}
\end{figure}

Figure~\ref{fig:qcd_power} shows the performance and energy consumption applying the power knobs described in Sec.~\ref{subsec:spmv_power} to the DW kernel.
For the highest performance setting ($\mathrm{eco}=0$ and $f=2.2\,\GHZ$) we see that the energy consumption of the RIRI implementation is almost $20$\% higher than for RRII.
Furthermore, the energy consumption of RRII can be reduced further without significant loss in performance by lowering the clock frequency and switching on eco mode.
However, for RIRI the performance drops drastically when eco mode is activated since here the bottleneck is the FLA pipeline (see Sec.~\ref{subsec:qcd_ecm_pred}).
Eco mode turns off the FLB pipeline, which increases the load on FLA and hence the runtime.
Regardless of the implementation, the power consumption of the DW kernel is about 205\,\W\ at the highest power and frequency settings. It reduces to about 145\,\W\ with $\mathrm{eco}=2$ and $f=2.0\,\GHZ$.

\subsection{Comparison with other architectures}

\begin{figure}[tb]
	\centering
	%
	\input{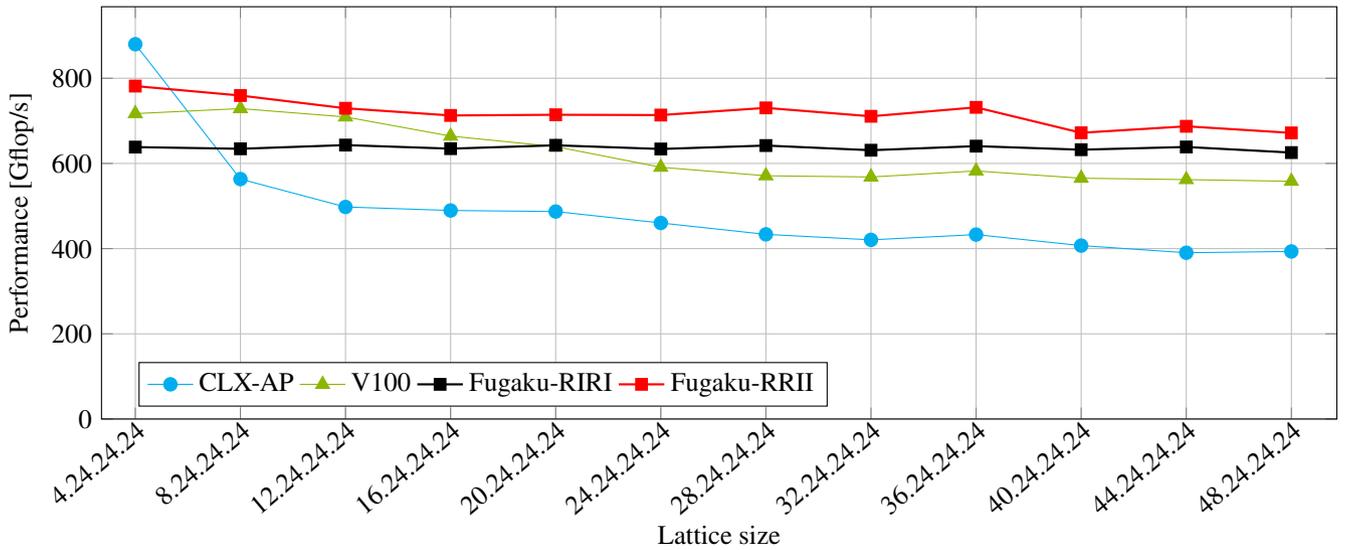}%

	\caption{Performance comparison of the DW kernel on Fugaku, Intel Cascade Lake AP and
  NVIDIA V100 for various lattice sizes with $L_s=8$.}
	\label{fig:qcd_arch_comparison}
\end{figure}

\label{subsec:qcd_arch_comparison}
In this section we briefly compare the DW kernel performance among the \afx, an Intel Cascade Lake CPU (CLX-AP) and a V100 GPU.
Figure~\ref{fig:qcd_arch_comparison} shows this comparison for various lattice sizes. The size of
the lattice is varied only along the innermost dimension~$x$.
The experiments on the V100 were conducted using the Grid Lattice QCD framework. We used our own implementations~\cite{Meyer:gridbench}
on the other architectures.
For CLX-AP and V100 we used the RIRI layout. This layout performs best on CLX-AP. For the V100 this
is the only layout currently available in Grid.
The results are qualitatively similar to that of the SpMV  performance comparison shown in Sec.~\ref{subsec:spmv_arch_comparison} as both  kernels are bandwidth bound. 
For smaller lattice sizes the CLX-AP has the advantage of large caches,
while for larger lattice sizes both the \afx\ and V100
have a $1.5\times$ performance advantage due to the higher memory bandwidth.
The \afx\ has a slight performance advantage over the
V100, however, we did not do a detailed inspection and analysis of the V100 code to see if there are any optimization opportunities.

\subsection{Further optimization options}
\label{sec:qcd_future}
The LC analysis suggests further optimization opportunities.
For example, cache blocking minimizes LC violations and facilitates data reuse in the caches.
Thread scheduling techniques increase data sharing between cores.
Zero fill instructions could be used to reduce the data traffic by avoiding write-allocate transfers.
Another potential improvement is to first
cut the inner dimensions ($s$ and $x$) for vectorization rather than the current approach of cutting outermost dimensions.
This will further relax the LC. 
However, these optimizations involve architecture-specific tuning parameters and code paths and thus impact code maintainability. 

We have not yet investigated the use of single precision,
which is often sufficient depending on the algorithm in which the
kernel is embedded.
This will be the subject of future work.

\section{Conclusions}\label{sec:summary}

\subsection{Summary and outlook}
We have further improved the ECM machine model for the \afx\ CPU introduced in \cite{PMBS20_A64FX} and showed its applicability to the Fugaku processor.
We validated the model with simple streaming kernels and could observe a high accuracy for in-memory data sets.
The memory hierarchy \rAdd{is}
partially overlapping, allowing for a substantial single-core memory bandwidth with optimized code.
Long floating-point instruction latencies and limited out-of-order execution capabilities were identified as the main culprits of poor performance and lack of bandwidth saturation.
Vector intrinsics and manual unrolling are often required to achieve high performance.

Further, we applied the ECM model to sparse matrix-vector multiplication to 
\rAdd{identify}
the impact of A64FX proprietary performance features like the sector cache and power-related tuning knobs.
The \sellcs\ matrix storage was shown to achieve performance and memory-bandwidth saturation 
superior to the standard CRS format. 
A comparison with state-of-the-art CPU and GPU architectures showed that for large, memory-bound \spmv\ datasets, the A64FX can outperform Intel's CLX-AP by a factor of two and is on par with NVIDIA's V100.

Finally, we used the ECM model for a comprehensive analysis of the Lattice QCD domain wall kernel in a subset of the Grid framework.
We showed that both compilers used for this work (FCC and GCC) exhibit a lack of quality in code optimization. A change of the data layout can achieve a better balance of the port pressure and
thus increase the performance significantly.
When comparing the energy consumption of data layouts, the split RRII data layout proved to be almost 20\% more energy efficient without a noticeable loss in performance.
The \afx\ shows a speedup of $1.5\times$ over the CLX-AP for problem sizes exceeding both architectures' last-level cache and comparable performance to a V100.

The present work opens up many interesting opportunities for future research.
For example, load-balancing issues together with the ccNUMA characteristics
of the \afx\ warrant further investigation.
The insight gained from the \spmv\ analysis can be applied to applications
such as Chebyshev filter diagonalization or linear solvers.
 Further optimizations to the QCD kernels have already been
 discussed in Sec.~\ref{sec:qcd_future}. 
 Last but not least a more detailed analysis of the mechanisms for power tuning
 is of great interest to all applications.

\subsection{Related work}
Since the \afx\ CPU is a very recent design, the amount of performance-centric research is limited.
Dongarra~\cite{Dongarra20Fugaku} reported on basic architectural features, HPC benchmarks (HPL, HPCG, HPL-AI) and the software environment of the Fugaku system.
Poenaru and McIntosh-Smith~\cite{Poenaru2020SVE} presented results on the effect of using wide vector registers and compared the performance and cache behavior of the A64FX for HPC benchmarks to the ThunderX2 platform.
Both Odajima et al.~\cite{Odajima2020perf} and Jackson et al.~\cite{jackson2020investigating} investigated benchmarks, full applications and proxy apps in comparison to Intel and other Arm-based systems but did not use performance models for analysis.
Gupta et al.~\cite{Gupta_a64fx} and Brank et al.~\cite{Brank_a64fx}  investigated stencil codes, proxy applications, \spmv\ and 
memory-bound fluid solvers on several Arm-based platforms including \afx\ but did not provide detailed 
and validated performance models.
Heybrock et al.~\cite{Heybrock_qcd} and Jo{\'o} et al.~\cite{Joo15} analyzed data traffic properties and data layout options for  QCD kernels on the Intel Xeon Phi architecture.
Kodama et al.~\cite{energy_a64fx} studied a comprehensive set of power-tuning knobs for STREAM and DGEMM kernels on Fugaku.

\section*{Acknowledgments}
We thank Daniel Richtmann for providing us with the V100 benchmarks of the DW kernel
and Julian Hammer for useful discussions regarding cache modeling.
This work used computational resources of the supercomputer   Fugaku provided by RIKEN
through the HPCI System Research Project (Project ID: hp200261).
We are indebted to HLRN for providing access to their Lise cluster.
This work was supported in part by KONWIHR, by DFG in the framework of SFB/TRR 55 and
by MEXT as ``Program for Promoting Researches on the
Supercomputer Fugaku'' (Simulation for basic science: from fundamental laws of
particles to creation of nuclei).

\section*{Disclaimer}
The results obtained on the evaluation environment in the trial phase do not
guarantee the performance, power and other attributes of the supercomputer Fugaku
at the start of its public use operation.

\bibliographystyle{WileyNJD-AMA}
\bibliography{references}%

\begin{thebibliography}{10}
\providecommand \doibase [0]{http://dx.doi.org/}%

\bibitem{PMBS20_A64FX}
{Alappat} C, {Laukemann} J, {Gruber} T, et al. Performance Modeling of
  Streaming Kernels and Sparse Matrix-Vector Multiplication on A64FX. In:
  Institute of Electrical and Electronics Engineers. ; 2020\string: 1-7.
\newblock \href{https://doi.org/10.1109/PMBS51919.2020.00006}{doi:
  10.1109/PMBS51919.2020.00006}

\bibitem{ACLE}
{Arm} . {ARM C Language Extensions for SVE}.
  \url{https://developer.arm.com/documentation/100987/0000/}; .
\newblock Accessed 2020-09-28.

\bibitem{ibench}
Hofmann J. {ibench -- Measure Instruction Latency and Throughput}.
  \url{https://github.com/RRZE-HPC/ibench};  2018.

\bibitem{a64fx_manual}
{Fujitsu Limited} . {\it {A64FX Microarchitecture Manual 1.3}}.
\newblock Fujitsu Limited .
\newblock 2020.
\newblock
  \url{https://github.com/fujitsu/A64FX/blob/1b3071af0369ee02b1752b7556e949050349985d/doc/A64FX_Microarchitecture_Manual_en_1.3.pdf}.

\bibitem{Treibig:2010:2}
Treibig J, Hager G, Wellein G. {LIKWID: A Lightweight Performance-Oriented Tool
  Suite for x86 Multicore Environments}. In: International Association for
  Computers and Communications. ; 2010\string: 207-216.
\newblock \href{http://dx.doi.org/10.1109/ICPPW.2010.38}{doi:
  10.1109/ICPPW.2010.38}

\bibitem{gruber_thomas_2020_4282696}
Gruber T, Eitzinger J, Hager G, Wellein G. RRZE-HPC/likwid: likwid-5.1.0.
  \url{https://github.com/RRZE-HPC/likwid/};  2020

\bibitem{PowerAPI}
{Grant} RE, {Levenhagen} M, {Olivier} SL, {DeBonis} D, {Pedretti} KT, {Laros
  III} JH. Standardizing Power Monitoring and Control at Exascale. {\it
  Computer} 2016\string; 49(10)\string: 38-46.
\newblock \href {\doibase 10.1109/MC.2016.308} {doi: 10.1109/MC.2016.308}

\bibitem{OSACA2018}
{Laukemann} J, {Hammer} J, {Hofmann} J, {Hager} G, {Wellein} G. {Automated
  Instruction Stream Throughput Prediction for Intel and AMD
  Microarchitectures}. In: Institute of Electrical and Electronics Engineers. ;
  2018\string: 121-131.
\newblock \href{http://dx.doi.org/10.1109/PMBS.2018.8641578}{doi:
  10.1109/PMBS.2018.8641578}.

\bibitem{OSACA2019}
Laukemann J, Hammer J, Hager G, Wellein G. Automatic Throughput and Critical
  Path Analysis of x86 and {ARM} Assembly Kernels. In: Institute of Electrical
  and Electronics Engineers. ; 2019\string: 1-6.
\newblock \href{http://dx.doi.org/10.1109/PMBS49563.2019.00006}{doi:
  10.1109/PMBS49563.2019.00006}

\bibitem{JSFI310}
Hofmann J, Alappat C, Hager G, Fey D, Wellein G. Bridging the Architecture Gap:
  Abstracting Performance-Relevant Properties of Modern Server Processors. {\it
  Supercomputing Frontiers and Innovations} 2020\string; 7(2).
\newblock \href {\doibase 10.14529/jsfi200204} {doi: 10.14529/jsfi200204}

\bibitem{sthw15}
Stengel H, Treibig J, Hager G, Wellein G. {Q}uantifying {P}erformance
  {B}ottlenecks of {S}tencil {C}omputations using the
  {E}xecution-{C}ache-{M}emory model. In: ICS '15. Association for Computing
  Machinery. Association for Computing Machinery; 2015; New York, NY, USA.
\newblock \href{https://doi.org/10.1145/2751205.2751240}{doi:
  10.1145/2751205.2751240}

\bibitem{Gropp:1999}
Gropp WD, Kaushik DK, Keyes DE, Smith BF. Towards Realistic Performance Bounds
  for Implicit {CFD} Codes. In:  Keyes D, Periaux J, Ecer A, Satofuka N, Fox P.
  \kern-2pt, eds. {\it Parallel Computational Fluid Dynamics 1999}~Elsevier.
  ~Elsevier; 2000\string: 241-248.
\newblock
  \url{https://wgropp.cs.illinois.edu/bib/papers/pdata/1999/pcfd99/gkks.ps}

\bibitem{Kreutzer14}
Kreutzer M, Hager G, Wellein G, Fehske H, Bishop AR. A Unified Sparse Matrix
  Data Format for Efficient General Sparse Matrix-Vector Multiplication on
  Modern Processors with Wide {SIMD} Units. {\it SIAM J. Sci. Comput.}
  2014\string; 36(5)\string: C401-C423.
\newblock \href {\doibase 10.1137/130930352} {doi: 10.1137/130930352}

\bibitem{MVE}
Lam M. Software Pipelining: An Effective Scheduling Technique for VLIW
  Machines. In: PLDI '88. Association for Computing Machinery. Association for
  Computing Machinery; 1988; New York, NY, USA\string: 318–328.
\newblock \href{https://doi.org/10.1145/53990.54022}{doi: 10.1145/53990.54022}

\bibitem{UOF}
Davis TA, Hu Y. The University of Florida Sparse Matrix Collection. {\it ACM
  Trans. Math. Softw.} 2011\string; 38(1)\string: 1:1--1:25.
\newblock \href {\doibase 10.1145/2049662.2049663} {doi:
  10.1145/2049662.2049663}

\bibitem{Kreutzer17}
Kreutzer M, Thies J, R{\"o}hrig-Z{\"o}llner M, et al. {GHOST}: Building Blocks
  for High Performance Sparse Linear Algebra on Heterogeneous Systems. {\it
  International Journal of Parallel Programming} 2017\string; 45\string:
  1046-1072.
\newblock \href {\doibase 10.1007/s10766-016-0464-z} {doi:
  10.1007/s10766-016-0464-z}

\bibitem{KAPLAN1992342}
Kaplan DB. A method for simulating chiral fermions on the lattice. {\it Physics
  Letters B} 1992\string; 288(3)\string: 342-347.
\newblock \href {\doibase 10.1016/0370-2693(92)91112-M} {doi:
  10.1016/0370-2693(92)91112-M}

\bibitem{Furman:1994ky}
Furman V, Shamir Y. {Axial symmetries in lattice QCD with Kaplan fermions}.
  {\it Nucl. Phys. B} 1995\string; 439\string: 54--78.
\newblock \href {\doibase 10.1016/0550-3213(95)00031-M} {doi:
  10.1016/0550-3213(95)00031-M}

\bibitem{Boyle:2015tjk}
Boyle P, Yamaguchi A, Cossu G, Portelli A. {Grid: A next generation data
  parallel C++ QCD library}. {\it PoS (LATTICE 2015)} 2016\string: 023.
\newblock \href {\doibase 10.22323/1.251.0023} {doi: 10.22323/1.251.0023}

\bibitem{Meyer:cluster18}
Georg P, Meyer N, Pleiter D, Solbrig S, Wettig T. {SVE-enabling Lattice QCD
  Codes}. {\it {2018 IEEE International Conference on Cluster Computing
  (CLUSTER)}} {2018}\string: 623-628.
\newblock \href {\doibase 10.1109/CLUSTER.2018.00079} {doi:
  10.1109/CLUSTER.2018.00079}

\bibitem{Meyer:2019gbz}
Meyer N, Pleiter D, Solbrig S, Wettig T. {Lattice QCD on upcoming Arm
  architectures}. {\it PoS (LATTICE 2018)} 2019\string: {316}.
\newblock \href {\doibase 10.22323/1.334.0316} {doi: 10.22323/1.334.0316}

\bibitem{Meyer:aplat2020}
Georg P, Meyer N, Pleiter D, Solbrig S, Wettig T. {Lattice QCD on QPACE 4}.
  \url{https://conference-indico.kek.jp/event/113/contributions/2139/attachments/1391/1545/aplat2020\_lqcd\_on\_qpace4\_meyer\_v2.pdf};
  2020.

\bibitem{Boyle:gridbench}
Boyle P, Yamaguchi A. {GridBench -- Single CPU benchmarks cutting down Grid}.
  \url{https://github.com/paboyle/GridBench};  2020.

\bibitem{Meyer:gridbench}
Meyer N, Alappat C. {GridBench -- AVX512 and A64FX extensions}.
  \url{https://github.com/nmeyer-ur/GridBench/tree/intrinsics};  2021.

\bibitem{INSPECT}
Hornich J, Hammer J, Hager G, Gruber T, Wellein G. Collecting and Presenting
  Reproducible Intranode Stencil Performance: INSPECT. {\it Supercomputing
  Frontiers and Innovations} 2019\string; 6(3).
\newblock \href {\doibase 10.14529/jsfi190301} {doi: 10.14529/jsfi190301}

\bibitem{Joo15}
Jo{\'o} B, Smelyanskiy M, Kalamkar DD, Vaidyanathan K. {{Chapter 9 - Wilson
  Dslash Kernel From Lattice QCD Optimization}}. In:  Reinders J, Jeffers J.
  \kern-2pt, eds. {\it High Performance Parallelism Pearls Volume Two:
  Multicore and Many-core Programming Approaches}. 2. Boston, MA, USA: Morgan
  Kaufmann.  2015 (pp. 139 - 170).
\newblock \href{http://dx.doi.org/10.1016/B978-0-12-803819-2.00023-9}{doi:
  10.1016/B978-0-12-803819-2.00023-9}

\bibitem{lc_origin}
Rivera G, Tseng CW. Tiling Optimizations for {3D} Scientific Computations. In:
  SC '00. IEEE Computer Society. IEEE Computer Society; 2000; USA\string:
  32--32.
\newblock \href{https://doi.org/10.1109/sc.2000.10015}{doi:
  10.1109/sc.2000.10015}

\bibitem{kerncraft}
Hammer J, Eitzinger J, Hager G, Wellein G. Kerncraft: A Tool for Analytic
  Performance Modeling of Loop Kernels. In:  Niethammer C, Gracia J, Hilbrich
  T, Kn{\"u}pfer A, Resch MM, Nagel WE. \kern-2pt, eds. {\it Tools for High
  Performance Computing 2016}Springer International Publishing. Springer
  International Publishing; 2017; Cham\string: 1--22.
\newblock \href{https://doi.org/10.1007/978-3-319-56702-0\_1}{doi:
  10.1007/978-3-319-56702-0\_1}

\bibitem{yasksite}
Alappat CL, Seiferth J, Hager G, Korch M, Rauber T, Wellein G. {YaskSite}:
  {Stencil} Optimization Techniques Applied to Explicit {ODE} Methods on Modern
  Architectures. In: 2021 IEEE/ACM International Symposium on Code Generation
  and Optimization (CGO). ; 2021\string: 174-186.
\newblock \href{https://doi.org/10.1109/CGO51591.2021.9370316}{doi:
  10.1109/CGO51591.2021.9370316}

\bibitem{Dongarra20Fugaku}
Dongarra J. {Report on the Fujitsu Fugaku System}. Tech. Rep. ICL-UT-20-06,
  University of Tennessee, Dept. of Electrical Engineering and Computer
  Science, Innovative Computing Laboratory; University of Tennessee, Knoxville,
  Oak Ridge National Laboratory:   2020.
\newblock
  \url{https://www.icl.utk.edu/files/publications/2020/icl-utk-1379-2020.pdf}.

\bibitem{Poenaru2020SVE}
{Poenaru} A, {McIntosh-Smith} S. The Effects of Wide Vector Operations on
  Processor Caches. In: Institute of Electrical and Electronics Engineers. ;
  2020\string: 531-539.
\newblock \href{https://doi.org/10.1109/CLUSTER49012.2020.00076}{doi:
  10.1109/CLUSTER49012.2020.00076}

\bibitem{Odajima2020perf}
{Odajima} T, {Kodama} Y, {Tsuji} M, {Matsuda} M, {Maruyama} Y, {Sato} M.
  Preliminary Performance Evaluation of the {F}ujitsu {A64FX} Using {HPC}
  Applications. In: Institute of Electrical and Electronics Engineers. ;
  2020\string: 523-530.
\newblock \href{https://doi.org/10.1109/CLUSTER49012.2020.00075}{doi:
  10.1109/CLUSTER49012.2020.00075}

\bibitem{jackson2020investigating}
{Jackson} A, {Weiland} M, {Brown} N, {Turner} A, {Parsons} M. Investigating
  Applications on the A64FX. In: Institute of Electrical and Electronics
  Engineers. ; 2020\string: 549-558.
\newblock \href{https://doi.org/10.1109/CLUSTER49012.2020.00078}{doi:
  10.1109/CLUSTER49012.2020.00078}

\bibitem{Gupta_a64fx}
{Gupta} N, {Ashiwal} R, {Brank} B, {Peddoju} SK, {Pleiter} D. Performance
  Evaluation of ParalleX Execution model on Arm-based Platforms. In: Institute
  of Electrical and Electronics Engineers. ; 2020\string: 567-575.
\newblock \href{https://doi.org/10.1109/CLUSTER49012.2020.00080}{doi:
  10.1109/CLUSTER49012.2020.00080}

\bibitem{Brank_a64fx}
{Brank} B, {Nassyr} S, {Pouyan} F, {Pleiter} D. Porting Applications to
  Arm-based Processors. In: Institute of Electrical and Electronics Engineers.
  ; 2020\string: 559-566.
\newblock \href{https://doi.org/10.1109/CLUSTER49012.2020.00079}{doi:
  10.1109/CLUSTER49012.2020.00079}

\bibitem{Heybrock_qcd}
Heybrock S, Jo\'{o} B, Kalamkar DD, et al. Lattice QCD with Domain
  Decomposition on Intel® Xeon Phi™ Co-Processors. In: SC '14. Institute of
  Electrical and Electronics Engineers. IEEE Press; 2014\string: 69–80.
\newblock \href{https://doi.org/10.1109/SC.2014.11}{doi: 10.1109/SC.2014.11}

\bibitem{energy_a64fx}
{Kodama} Y, {Odajima} T, {Arima} E, {Sato} M. Evaluation of Power Management
  Control on the Supercomputer Fugaku. In: Institute of Electrical and
  Electronics Engineers. ; 2020\string: 484-493.
\newblock \href{https://doi.org/10.1109/CLUSTER49012.2020.00069}{doi:
  10.1109/CLUSTER49012.2020.00069}

\end{thebibliography}

\end{document}